\documentclass[twocolumn, prx, aps, superscriptaddress, floatfix, nofootinbib,10pt]{revtex4-2}

\usepackage[utf8]{inputenc}
\usepackage[english]{babel}
\usepackage[sort&compress]{natbib}
\usepackage[a4paper, total={170mm, 237mm}]{geometry}

\usepackage{graphicx}
\usepackage{amsmath, amssymb, amsthm}
\usepackage[normalem]{ulem}
\usepackage{tensor}
\usepackage{braket}
\usepackage{cases}
\usepackage{dsfont}
\usepackage[dvipsnames]{xcolor}
\usepackage{hyperref}
\usepackage{nameref}
\hypersetup{
    colorlinks = true,
    allcolors = RoyalPurple
}
\usepackage[percent]{overpic}
\usepackage{svg}

\usepackage{tikz}
\usepackage{circuitikz}
\usetikzlibrary{shapes}

\usepackage[export]{adjustbox}
\usepackage{leftidx}
\usepackage{BOONDOX-cal} 
\usepackage{orcidlink}
\usepackage{complexity}

\DeclareMathOperator{\Mat}{Mat}
\DeclareMathOperator{\Tr}{Tr}

\newtheorem{definition}{Definition}[]

\newtheorem{lemma}[definition]{Lemma}

\newcommand{\comm}[2]{\left[ #1 , #2 \right]}

\renewcommand{\ket}[1]{\vert #1 \rangle}
\renewcommand{\bra}[1]{\langle #1 \vert}

\newcommand{\norm}[1]{\left\lVert #1 \right\rVert}

\renewcommand{\epsilon}{\varepsilon}

\definecolor{BLUE}{RGB}{0,0,255}

\newcommand{\costgradpreI}{
\begin{tikzpicture}
    \draw (-1.6,0) -- (1.5,0)
    (-0.5,0) -- (-0.5,0.5)
    (-2,0.5) -- (-0.5,0.5)
    (-2,0.5) -- (-2,-3.5)
    (-0.5,-3.5) -- (-2,-3.5)
    (0.5,0) -- (0.5,-1)
    ;
    \filldraw[fill=Lavender] (0.3, -0.25) rectangle node {$\sigma_r$} (0.7, -0.65);
    \filldraw[fill=Lavender] (-1.8, -1.3) rectangle node {$\sigma_l$} (-2.2, -1.7);
    \draw node at (-0.35,0.3) {\small $\Gamma_l$};
    \draw node at (0.5,0.3) {\small $\Gamma_r$};
    \filldraw[fill=LimeGreen] (0,0) circle (0.25cm) node{$\Lambda_0$};
    \filldraw[fill=LimeGreen] (-1,0) circle (0.25cm) node{$\Lambda_l$};
    \filldraw[fill=LimeGreen] (1,0) circle (0.25cm) node{$\Lambda_r$};
    \draw (-1.5,-1) -- (1.5,-1)
    (-0.5,-1) -- (-0.5,-2)
    (-1.5,-2) -- (-1.5,-1)
    (1.5,-1) -- (1.5,0)
    ;
    \draw node at (-0.5,-0.65) {\small$\bar\Gamma_l$};
    \draw node at (0.5,-1.325) {\small$\bar\Gamma_r$};
    \filldraw[fill=LimeGreen] (0,-1) circle (0.25cm) node{$\Lambda_0$};
    \filldraw[fill=LimeGreen] (-1,-1) circle (0.25cm) node{$\Lambda_l$};
    \filldraw[fill=LimeGreen] (1,-1) circle (0.25cm) node{$\Lambda_r$};
    \draw (-1.5,-2) -- (1.5,-2)
    (0.5,-2) -- (0.5,-3)
    ;
    \filldraw[fill=LimeGreen] (0,-2) circle (0.25cm) node{$\Lambda_0$};
    \filldraw[fill=LimeGreen] (-1,-2) circle (0.25cm) node{$\Lambda_l$};
    \filldraw[fill=LimeGreen] (1,-2) circle (0.25cm) node{$\Lambda_r$};
    \draw node at (-0.5,-2.3) {\small$\Gamma_l$};
    \draw node at (0.5,-1.7) {\small$\Gamma_r$};
    \draw (-1.6,-3) -- (1.5,-3)
    (-0.5,-3) -- (-0.5,-3.5)
    (-1.6,-3) -- (-1.6,0)
    (1.5,-3) -- (1.5,-2)
    ;
    \filldraw[fill=LimeGreen] (0,-3) circle (0.25cm) node{$\Lambda_0$};
    \filldraw[fill=LimeGreen] (-1,-3) circle (0.25cm) node{$\Lambda_l$};
    \filldraw[fill=LimeGreen] (1,-3) circle (0.25cm) node{$\Lambda_r$};
    \draw node at (-0.5,-2.7) {\small$\bar\Gamma_l$};
    \draw node at (0.5,-3.35) {\small$\bar\Gamma_r$};
    \filldraw[fill=SkyBlue!50!black] (-0.625,-1.125) rectangle (-0.375, -.875);
    \filldraw[fill=SkyBlue!50!black] (-0.625,-.125) rectangle (-0.375, .125);
    \filldraw[fill=SkyBlue!50!black] (0.375,-.125) rectangle (0.625, .125);
    \filldraw[fill=SkyBlue!50!black] (0.375,-1.125) rectangle (0.625, -.875);
    \filldraw[fill=SkyBlue!50!black] (-0.625,-2.125) rectangle (-0.375, -1.875);
    \filldraw[fill=SkyBlue!50!black] (-0.625,-3.125) rectangle (-0.375, -2.875);
    \filldraw[fill=SkyBlue!50!black] (0.375,-3.125) rectangle (0.625, -2.875);
    \filldraw[fill=SkyBlue!50!black] (0.375,-2.125) rectangle (0.625, -1.875);
\end{tikzpicture}
}

\newcommand{\costgradpreII}{
\begin{tikzpicture}
    \draw (-1.6,0) -- (1.5,0)
    (-0.5,0) -- (-0.5,0.5)
    (-2,0.5) -- (-0.5,0.5)
    (-2,0.5) -- (-2,-3.5)
    (-0.5,-3.5) -- (-2,-3.5)
    (0.5,0) -- (0.5,-1)
    ;
    \filldraw[fill=Lavender] (0.3, -0.25) rectangle node {$\sigma_r$} (0.7, -0.65);
    \draw node at (-0.35,0.25) {\small$\Gamma_l$};
    \draw node at (0.5,0.3) {\small$\Gamma_r$};
    \filldraw[fill=LimeGreen] (0,0) circle (0.25cm) node{$\Lambda_0$};
    \filldraw[fill=LimeGreen] (-1,0) circle (0.25cm) node{$\Lambda_l$};
    \filldraw[fill=LimeGreen] (1,0) circle (0.25cm) node{$\Lambda_r$};
    \draw (-1.5,-1) -- (1.5,-1)
    (-0.5,-1) -- (-0.5,-2)
    (-1.5,-1) -- (-1.5,-2)
    (1.5,-1) -- (1.5,0)
    ;
    \draw node at (-0.5,-0.7) {\small$\bar\Gamma_l$};
    \draw node at (0.5,-1.325) {\small$\bar\Gamma_r$};
    \filldraw[fill=LimeGreen] (0,-1) circle (0.25cm) node{$\Lambda_0$};
    \filldraw[fill=LimeGreen] (-1,-1) circle (0.25cm) node{$\Lambda_l$};
    \filldraw[fill=LimeGreen] (1,-1) circle (0.25cm) node{$\Lambda_r$};
    \draw (-1.5,-2) -- (1.5,-2)
    (0.5,-2) -- (0.5,-3)
    ;
    \filldraw[fill=LimeGreen] (0,-2) circle (0.25cm) node{$\Lambda_0$};
    \filldraw[fill=LimeGreen] (-1,-2) circle (0.25cm) node{$\Lambda_l$};
    \filldraw[fill=LimeGreen] (1,-2) circle (0.25cm) node{$\Lambda_r$};
    \draw node at (-0.5,-2.3) {\small$\Gamma_l$};
    \draw node at (0.5,-1.7) {\small$\Gamma_r$};
    \draw (-1.6,-3) -- (1.5,-3)
    (-0.5,-3) -- (-0.5,-3.5)
    (-1.6,-3) -- (-1.6,0)
    (1.5,-3) -- (1.5,-2)
    ;
    \filldraw[fill=Lavender] (-0.7, -1.7) rectangle node {$\sigma_l$} (-0.3, -1.3);
    \filldraw[fill=LimeGreen] (0,-3) circle (0.25cm) node{$\Lambda_0$};
    \filldraw[fill=LimeGreen] (-1,-3) circle (0.25cm) node{$\Lambda_l$};
    \filldraw[fill=LimeGreen] (1,-3) circle (0.25cm) node{$\Lambda_r$};
    \draw node at (-0.5,-2.7) {\small$\bar\Gamma_l$};
    \draw node at (0.5,-3.35) {\small$\bar\Gamma_r$};
    \filldraw[fill=SkyBlue!50!black] (-0.625,-1.125) rectangle (-0.375, -.875);
    \filldraw[fill=SkyBlue!50!black] (-0.625,-.125) rectangle (-0.375, .125);
    \filldraw[fill=SkyBlue!50!black] (0.375,-.125) rectangle (0.625, .125);
    \filldraw[fill=SkyBlue!50!black] (0.375,-1.125) rectangle (0.625, -.875);
    \filldraw[fill=SkyBlue!50!black] (-0.625,-2.125) rectangle (-0.375, -1.875);
    \filldraw[fill=SkyBlue!50!black] (-0.625,-3.125) rectangle (-0.375, -2.875);
    \filldraw[fill=SkyBlue!50!black] (0.375,-3.125) rectangle (0.625, -2.875);
    \filldraw[fill=SkyBlue!50!black] (0.375,-2.125) rectangle (0.625, -1.875);
\end{tikzpicture}
}

\newcommand{\costgradpostI}{
\begin{tikzpicture}
    \draw (-1.5,0) -- (1.5,0)
    (-0.5,0) -- (-0.5,-1)
    (0.5,0) -- (0.5,-1)
    ;
    \filldraw[fill=Lavender] (0.3, -0.25) rectangle node {$\sigma_r$} (0.7, -0.65);
    \filldraw[fill=Lavender] (-0.3, -0.25) rectangle node {$\sigma_l$} (-0.7, -0.65);
    \draw node at (-0.5,0.3) {$\Gamma_l$};
    \draw node at (0.5,0.3) {$\Gamma_r$};
    \filldraw[fill=LimeGreen] (0,0) circle (0.25cm) node{$\Lambda_0$};
    \filldraw[fill=LimeGreen] (-1,0) circle (0.25cm) node{$\Lambda_l$};
    \filldraw[fill=LimeGreen] (1,0) circle (0.25cm) node{$\Lambda_r$};
    \draw (-1.5,-1) -- (1.5,-1)
    (-1.5,-1) -- (-1.5,0)
    (1.5,-1) -- (1.5,0)
    ;
    \draw node at (-0.5,-1.35) {$\bar\Gamma_l$};
    \draw node at (0.5,-1.35) {$\bar\Gamma_r$};
    \filldraw[fill=LimeGreen] (0,-1) circle (0.3cm) node{$\Lambda_0^3$};
    \filldraw[fill=LimeGreen] (-1,-1) circle (0.25cm) node{$\Lambda_l$};
    \filldraw[fill=LimeGreen] (1,-1) circle (0.25cm) node{$\Lambda_r$};
    \filldraw[fill=SkyBlue!50!black] (-0.625,-1.125) rectangle (-0.375, -.875);
    \filldraw[fill=SkyBlue!50!black] (-0.625,-.125) rectangle (-0.375, .125);
    \filldraw[fill=SkyBlue!50!black] (0.375,-.125) rectangle (0.625, .125);
    \filldraw[fill=SkyBlue!50!black] (0.375,-1.125) rectangle (0.625, -.875);
\end{tikzpicture}
}

\newcommand{\costgradpostII}{
\begin{tikzpicture}
    \draw (-1.5,0) -- (1.5,0)
    (-0.5,0) -- (-0.5,-1)
    (0.5,0) -- (0.5,-1)
    ;
    \filldraw[fill=Lavender] (0.3, -0.25) rectangle node {$\sigma_r$} (0.7, -0.65);
    \filldraw[fill=Lavender] (-0.3, -0.25) rectangle node {$\sigma_l$} (-0.7, -0.65);
    \draw node at (-0.5,0.3) {$\Gamma_l$};
    \draw node at (0.5,0.3) {$\Gamma_r$};
    \filldraw[fill=LimeGreen] (0,0) circle (0.3cm) node{$\Lambda_0^3$};
    \filldraw[fill=LimeGreen] (-1,0) circle (0.25cm) node{$\Lambda_l$};
    \filldraw[fill=LimeGreen] (1,0) circle (0.25cm) node{$\Lambda_r$};
    \draw (-1.5,-1) -- (1.5,-1)
    (-1.5,-1) -- (-1.5,0)
    (1.5,-1) -- (1.5,0)
    ;
    \draw node at (-0.5,-1.35) {$\bar\Gamma_l$};
    \draw node at (0.5,-1.35) {$\bar\Gamma_r$};
    \filldraw[fill=LimeGreen] (0,-1) circle (0.25cm) node{$\Lambda_0$};
    \filldraw[fill=LimeGreen] (-1,-1) circle (0.25cm) node{$\Lambda_l$};
    \filldraw[fill=LimeGreen] (1,-1) circle (0.25cm) node{$\Lambda_r$};
    \filldraw[fill=SkyBlue!50!black] (-0.625,-1.125) rectangle (-0.375, -.875);
    \filldraw[fill=SkyBlue!50!black] (-0.625,-.125) rectangle (-0.375, .125);
    \filldraw[fill=SkyBlue!50!black] (0.375,-.125) rectangle (0.625, .125);
    \filldraw[fill=SkyBlue!50!black] (0.375,-1.125) rectangle (0.625, -.875);
\end{tikzpicture}
}

\newcommand{\MPSGL}{
\begin{tikzpicture}
    \draw (-3,-1) -- (-0.5,-1)
    (-3,-0.5) -- (-3,-1)
    (-2,-0.5) -- (-2,-1)
    (-1,-0.5) -- (-1,-1)
    (1,-0.5) -- (1,-1)
    (2,-0.5) -- (2,-1)
    (0.5, -1) -- (2, -1)
    ;
    \draw node at (-3,-1.5) {$\Gamma_1$};
    \draw node at (-2,-1.5) {$\Gamma_2$};
    \draw node at (1,-1.5) {$\Gamma_{n-1}$};
    \draw node at (2,-1.5) {$\Gamma_n$};
    \filldraw[fill=LimeGreen] (-2.5,-1) circle (0.25cm) node{$\Lambda_1$};
    \filldraw[fill=LimeGreen] (-1.5,-1) circle (0.25cm) node{$\Lambda_2$};
    \draw node at (0,-1) {$\dots$};
    
    \filldraw[fill=LimeGreen] (1.5,-1) circle (0.25cm) node {};
    \draw node at (1.5,-0.5) {$\Lambda_{n-1}$};
    \filldraw[fill=SkyBlue!50!black] (-3.125,-1.125) rectangle (-2.875, -.875);
    \filldraw[fill=SkyBlue!50!black] (-2.125,-1.125) rectangle (-1.875, -.875);
    \filldraw[fill=SkyBlue!50!black] (-1.125,-1.125) rectangle (-.875, -.875);
    \filldraw[fill=SkyBlue!50!black] (.875,-1.125) rectangle (1.125, -.875);
    \filldraw[fill=SkyBlue!50!black] (1.875,-1.125) rectangle (2.125, -.875);
\end{tikzpicture}
}

\newcommand{\gateActionI}{
\begin{tikzpicture}
    \draw (-3,0) -- (3,0)
    (-1,0) -- (-1,1.75)
    (1,0) -- (1,1.75)
    ;
    \draw node at (-1,-0.4) {$\Gamma_l$};
    \draw node at (1,-0.4) {$\Gamma_r$};
    \filldraw[fill=LimeGreen] (-2,0) circle (0.35cm) node{$\Lambda_l$};
    \filldraw[fill=LimeGreen] (2,0) circle (0.35cm) node{$\Lambda_r$};
    \filldraw[fill=LimeGreen] (0,0) circle (0.35cm) node{$\Lambda_0$};
    \filldraw[fill=NavyBlue] (-1.5, 0.75) rectangle node {\color{white} $\mathbf{G_{l,r}}$} (1.5, 1.25);
    \filldraw[fill=SkyBlue!50!black] (-.875,-.125) rectangle (-1.125, .125);
    \filldraw[fill=SkyBlue!50!black] (.875,-.125) rectangle (1.125, .125);
\end{tikzpicture}
}

\newcommand{\gateActionII}{
\begin{tikzpicture}
    \draw (-3,0) -- (3,0)
    (-1,0) -- (-1,1.75)
    (1,0) -- (1,1.75)
    (-1,1) -- (1,1)
    ;
    \draw node at (-1,-0.4) {$\Gamma_l$};
    \draw node at (1,-0.4) {$\Gamma_r$};
    \draw node at (0,1.25) {$D \leq 4$};
    \filldraw[fill=LimeGreen] (-2,0) circle (0.35cm) node {$\Lambda_l$};
    \filldraw[fill=LimeGreen] (2,0) circle (0.35cm) node {$\Lambda_r$};
    \filldraw[fill=LimeGreen] (0,0) circle (0.35cm) node {$\Lambda_0$};
    \filldraw[fill=NavyBlue] (-1, 1) circle (0.35cm);
    \filldraw[fill=NavyBlue] (1, 1) circle (0.35cm);
    \filldraw[fill=SkyBlue!50!black] (-.875,-.125) rectangle (-1.125, .125);
    \filldraw[fill=SkyBlue!50!black] (.875,-.125) rectangle (1.125, .125);
\end{tikzpicture}
}

\newcommand{\gateActionIV}{
\begin{tikzpicture}
    \draw (-1.5,0) -- (1.5,0)
    (-1,0) -- (-1,0.5)
    (1,0) -- (1,0.5)
    ;
    \filldraw[fill=NavyBlue!50] (-1.25,-0.25) rectangle node {$U_l \Lambda'_0 V_r$} (1.25, 0.25);
\end{tikzpicture}
}

\newcommand{\gateActionV}{
\begin{tikzpicture}
    \draw (-3,0) -- (3,0)
    (-1,0) -- (-1,0.5)
    (1,0) -- (1,0.5)
    ;
    \draw[white] (0,0) -- (0,1.5);
    \draw node at (-1,-0.4) {$\Gamma_l'$};
    \draw node at (1,-0.4) {$\Gamma_r'$};
    \filldraw[fill=LimeGreen] (-2,0) circle (0.35cm) node {$\Lambda_l$};
    \filldraw[fill=LimeGreen] (2,0) circle (0.35cm) node {$\Lambda_r$};
    \filldraw[fill=NavyBlue!50] (0,0) circle (0.35cm) node {$\Lambda_0'$};
    \filldraw[fill=SkyBlue!50!black] (-.875,-.125) rectangle (-1.125, .125);
    \filldraw[fill=SkyBlue!50!black] (.875,-.125) rectangle (1.125, .125);
\end{tikzpicture}
}

\newcommand{\disentangle}{
\begin{tikzpicture}
    \fill[fill=NavyBlue!30] (-4.5,0.175) rectangle (5.5, 3.875) node[anchor=north west] {\huge \color{NavyBlue} $U$};
    \draw (-4,0) -- (5,0)
    (-1,0) -- (-1,4)
    (1,0) -- (1,4)
    (-2,0) -- (-2,4)
    (2,0) -- (2,4)
    (-3,0) -- (-3,4)
    (3,0) -- (3,4)
    (0,0) -- (0,4)
    (-4,0) -- (-4,4)
    (4,0) -- (4,4)
    (5,0) -- (5,4)
    ;
    \draw node at (-3.5,-0.125) {\tiny $D$};
    \draw node at (-2.5,-0.125) {\tiny $D$};
    \draw node at (-1.5,-0.125) {\tiny $D$};
    \draw node at (-0.5,-0.125) {\tiny $D$};
    \draw node at (0.5,-0.125) {\tiny $D$};
    \draw node at (-4,-0.35) {$A_{[1]}$};
    \draw node at (-3,-0.35) {$A_{[2]}$};
    \draw node at (-2,-0.35) {$A_{[3]}$};
    \draw node at (-1,-0.35) {$A_{[4]}$};
    \draw node at (0,-0.35) {$A_{[5]}$};
    \draw node at (1,-0.35) {$A_{[6]}$};
    \draw node at (3,-0.35) {$\dots$};
    \draw node at (5,-0.35) {$A_{[n]}$};
    \filldraw[fill=NavyBlue] (-4.25,0.25) rectangle node {\color{white} $G_{1,2}$} (-2.75, 0.75);
    \filldraw[fill=NavyBlue] (-2.25,0.25) rectangle node {\color{white} $G_{3,4}$} (-0.75, 0.75);
    \filldraw[fill=NavyBlue] (-0.25,0.25) rectangle node {\color{white} $G_{5,6}$} (1.25, 0.75);
    \filldraw[fill=NavyBlue] (1.75,0.25) rectangle node {\color{white} $\dots$} (3.25, 0.75);
    \filldraw[fill=NavyBlue] (3.75,0.25) rectangle node {\color{white} $G_{n-1,n}$} (5.25, 0.75);
    \filldraw[fill=NavyBlue] (-3.25,1) rectangle node {\color{white} $G_{2,3}$} (-1.75, 1.5);
    \filldraw[fill=NavyBlue] (-1.25,1) rectangle node {\color{white} $G_{4,5}$} (0.25, 1.5);
    \filldraw[fill=NavyBlue] (0.75,1) rectangle node {\color{white} $\dots$} (2.25, 1.5);
    \filldraw[fill=NavyBlue] (2.75,1) rectangle node {\color{white} $G_{n-2, n-1}$} (4.25, 1.5);
    \filldraw[fill=NavyBlue] (-4.25,1.75) rectangle (-2.75, 2.25);
    \filldraw[fill=NavyBlue] (-2.25,1.75) rectangle (-0.75, 2.25);
    \filldraw[fill=NavyBlue] (-0.25,1.75) rectangle (1.25, 2.25);
    \filldraw[fill=NavyBlue] (1.75,1.75) rectangle (3.25, 2.25);
    \filldraw[fill=NavyBlue] (3.75,1.75) rectangle (5.25, 2.25);
    \filldraw[fill=NavyBlue] (-3.25,2.5) rectangle (-1.75, 3);
    \filldraw[fill=NavyBlue] (-1.25,2.5) rectangle (0.25, 3);
    \filldraw[fill=NavyBlue] (0.75,2.5) rectangle (2.25, 3);
    \filldraw[fill=NavyBlue] (2.75,2.5) rectangle (4.25, 3);
    \filldraw[fill=NavyBlue!70] (-4.25,3.25) rectangle (-3.75, 3.75);
    \filldraw[fill=NavyBlue!70] (-3.25,3.25) rectangle (-2.75, 3.75);
    \filldraw[fill=NavyBlue!70] (-2.25,3.25) rectangle (-1.75, 3.75);
    \filldraw[fill=NavyBlue!70] (-1.25,3.25) rectangle (-0.75, 3.75);
    \filldraw[fill=NavyBlue!70] (-0.25,3.25) rectangle (0.25, 3.75);
    \filldraw[fill=NavyBlue!70] (0.75,3.25) rectangle (1.25, 3.75);
    \filldraw[fill=NavyBlue!70] (1.75,3.25) rectangle (2.25, 3.75);
    \filldraw[fill=NavyBlue!70] (2.75,3.25) rectangle (3.25, 3.75);
    \filldraw[fill=NavyBlue!70] (3.75,3.25) rectangle (4.25, 3.75);
    \filldraw[fill=NavyBlue!70] (4.75,3.25) rectangle (5.25, 3.75);
    
    \filldraw[fill=SkyBlue!50!LimeGreen] (-3.875,-0.125) rectangle (-4.125, 0.125);
    \filldraw[fill=SkyBlue!50!LimeGreen] (-2.875,-0.125) rectangle (-3.125, 0.125);
    \filldraw[fill=SkyBlue!50!LimeGreen] (-1.875,-0.125) rectangle (-2.125, 0.125);
    \filldraw[fill=SkyBlue!50!LimeGreen] (-.875,-0.125) rectangle (-1.125, 0.125);
    \filldraw[fill=SkyBlue!50!LimeGreen] (.125,-0.125) rectangle (-.125, 0.125);
    \filldraw[fill=SkyBlue!50!LimeGreen] (1.125,-0.125) rectangle (.875, 0.125);
    \filldraw[fill=SkyBlue!50!LimeGreen] (2.125,-0.125) rectangle (1.875, 0.125);
    \filldraw[fill=SkyBlue!50!LimeGreen] (3.125,-0.125) rectangle (2.875, 0.125);
    \filldraw[fill=SkyBlue!50!LimeGreen] (3.875,-0.125) rectangle (4.125, 0.125);
    \filldraw[fill=SkyBlue!50!LimeGreen] (4.875,-0.125) rectangle (5.125, 0.125);
    \draw node at (6.5,2.5) {\huge $\approx$};
    \draw (8,2) -- (8,3)
    (9,2) -- (9,3)
    (10,2) -- (10,3)
    (11,2) -- (11,3)
    (12,2) -- (12,3)
    (11.5,2) -- (11.5,3)
    (10.5,2) -- (10.5,3)
    (9.5,2) -- (9.5,3)
    (8.5,2) -- (8.5,3)
    ;
    \filldraw[fill=SkyBlue!50!LimeGreen] (7.875,1.875) rectangle (8.125, 2.125);
    \filldraw[fill=SkyBlue!50!LimeGreen] (8.875,1.875) rectangle (9.125, 2.125);
    \filldraw[fill=SkyBlue!50!LimeGreen] (9.875,1.875) rectangle (10.125, 2.125);
    \filldraw[fill=SkyBlue!50!LimeGreen] (10.875,1.875) rectangle (11.125, 2.125);
    \filldraw[fill=SkyBlue!50!LimeGreen] (11.875,1.875) rectangle (12.125, 2.125);
    \filldraw[fill=SkyBlue!50!LimeGreen] (8.375,1.875) rectangle (8.625, 2.125);
    \filldraw[fill=SkyBlue!50!LimeGreen] (9.375,1.875) rectangle (9.625, 2.125);
    \filldraw[fill=SkyBlue!50!LimeGreen] (10.375,1.875) rectangle (10.625, 2.125);
    \filldraw[fill=SkyBlue!50!LimeGreen] (11.375,1.875) rectangle (11.625, 2.125);
    \draw node at (8.25,2.5) {$\otimes$};
    \draw node at (8.75,2.5) {$\otimes$};
    \draw node at (9.25,2.5) {$\otimes$};
    \draw node at (9.75,2.5) {$\otimes$};
    \draw node at (10.25,2.5) {$\otimes$};
    \draw node at (10.75,2.5) {$\otimes$};
    \draw node at (11.25,2.5) {$\otimes$};
    \draw node at (11.75,2.5) {$\otimes$};
    \draw node at (8,1.5) {$\ket{0}$};
    \draw node at (8.5,1.5) {$\ket{0}$};
    \draw node at (9,1.5) {$\ket{0}$};
    \draw node at (9.5,1.5) {$\ket{0}$};
    \draw node at (10,1.5) {$\ket{0}$};
    \draw node at (10.5,1.5) {$\ket{0}$};
    \draw node at (11,1.5) {$\ket{0}$};
    \draw node at (11.5,1.5) {$\ket{0}$};
    \draw node at (12,1.5) {$\ket{0}$};
    \draw node at (9,0.5) {\Large $\implies U^\dagger \ket{0}^{\otimes n} \approx \ket{\psi}$};
\end{tikzpicture}
}

\newcommand{\logicalbell}{
\begin{tikzpicture}[scale=0.8]
    \fill[fill=teal!30] (-5,0) -- (-0.5,0) -- (-0.625, 0.2) -- (-2.75,1) -- (-4.875,0.2);
    \fill[fill=teal!30] (4.5,0) -- (0,0) -- (0.125, 0.2) -- (2.25,1) -- (4.375, 0.2);
    \filldraw (-2.75,1) circle (0.2cm);
    \filldraw (2.25,1) circle (0.2cm);
    \draw[ultra thick] (-2.75,1) -- (2.25,1);
    \filldraw[Green] (-2.75, 0) circle (0.25cm)
    (-3.75, 0) circle (0.25cm)
    (-4.75, 0) circle (0.25cm)
    (-1.75, 0) circle (0.25cm)
    (-0.75, 0) circle (0.25cm)
    (0.25, 0) circle (0.25cm)
    (1.25, 0) circle (0.25cm)
    (2.25, 0) circle (0.25cm)
    (3.25, 0) circle (0.25cm)
    (4.25, 0) circle (0.25cm)
    ;
    \fill[fill=teal!30] (-5,-2.25) -- (-4.5,-2.25) -- (-0.25,-1.5) -- (-5, -2.25)
    (-3,-2.25) -- (-2.5,-2.25) -- (-0.25, -1.5) -- (-3, -2.25)
    (-1,-2.25) -- (-0.5,-2.25) -- (-0.25,-1.5) -- (-1, -2.25)
    (1,-2.25) -- (1.5,-2.25) -- (-0.25, -1.5) -- (1, -2.25)
    (3,-2.25) -- (3.5,-2.25) -- (-0.25,-1.5) -- (3, -2.25)
    ;
    \fill[fill=teal!30] (-4,-2.75) -- (-3.5,-2.75) -- (-0.25,-3.5) -- (-4, -2.75)
    (-2,-2.75) -- (-1.5,-2.75) -- (-0.25,-3.5) -- (-2, -2.75)
    (0,-2.75) -- (0.5,-2.75) -- (-0.25,-3.5) -- (0, -2.75)
    (2,-2.75) -- (2.5,-2.75) -- (-0.25,-3.5) -- (2, -2.75)
    (4,-2.75) -- (4.5,-2.75) -- (-0.25,-3.5) -- (4, -2.75)
    ;
    \filldraw (-0.25,-3.5) circle (0.2cm);
    \filldraw (-0.25,-1.5) circle (0.2cm);
    \draw[ultra thick] (-0.25,-3.5) -- (-0.25,-1.5) ;
    \filldraw[Green] (-2.75, -2.25) circle (0.25cm)
    (-3.75, -2.75) circle (0.25cm)
    (-4.75, -2.25) circle (0.25cm)
    (-1.75, -2.75) circle (0.25cm)
    (-0.75, -2.25) circle (0.25cm)
    (0.25, -2.75) circle (0.25cm)
    (1.25, -2.25) circle (0.25cm)
    (2.25, -2.75) circle (0.25cm)
    (3.25, -2.25) circle (0.25cm)
    (4.25, -2.75) circle (0.25cm)
    ;
\end{tikzpicture}
}

\newcommand{\AKLT}{
\begin{tikzpicture}
    \filldraw (-2.5,0) circle (0.1cm);
    \filldraw (-2,0) circle (0.1cm);
    \filldraw (-1,0) circle (0.1cm);
    \filldraw (-0.5,0) circle (0.1cm);
    \filldraw (0.5,0) circle (0.1cm);
    \filldraw (1,0) circle (0.1cm);
    \filldraw (2,0) circle (0.1cm);
    \filldraw (2.5,0) circle (0.1cm);
    \filldraw (3.5,0) circle (0.1cm);
    \filldraw (4,0) circle (0.1cm);
    \filldraw (5,0) circle (0.1cm);
    \filldraw (5.5,0) circle (0.1cm);
    \draw[thick] (-2,0) -- (-1,0)
    (-0.5,0) -- (0.5,0)
    (1,0) -- (2,0)
    (2.5,0) -- (3.5,0)
    (4,0) -- (5,0)
    ;
    \draw[] (-2.25,0) ellipse (0.6cm and 0.4cm);
    \draw[] (-.75,0) ellipse (0.6cm and 0.4cm);
    \draw[] (0.75,0) ellipse (0.6cm and 0.4cm);
    \draw[] (2.25,0) ellipse (0.6cm and 0.4cm);
    \draw[] (3.75,0) ellipse (0.6cm and 0.4cm);
    \draw[] (5.25,0) ellipse (0.6cm and 0.4cm);
\end{tikzpicture}
}

\usetikzlibrary{calc,shapes.geometric}
\newcommand{\isometry}[2]{
\begin{tikzpicture}
    \draw (-1,0) -- (0,0)
    (-0.5,0) -- (-0.5,0.5)
    ;
    \filldraw[fill=SkyBlue!50!LimeGreen] (-0.75, -0.25) rectangle node[black] {#1} (-0.25, 0.25);
    \draw node at (0.5,0) {$=:$};
    \draw (1,0) -- (2,0)
    (1.5,-0.5) -- (1.5,0.5)
    ;
    \filldraw[fill=NavyBlue!50] (1.25,-0.25) rectangle node[black] {#2} (1.75, 0.25);
    \filldraw[fill=gray] (1.7,-0.5) arc (0:-180:0.2)
    (1.7, -0.5) -- (1.3, -0.5)
    ;
    \draw node at (1.5, -1) {$\ket{0}$};
\end{tikzpicture}
}

\newcommand{\gradAB}[1][1]{
\begin{tikzpicture}[scale=#1]
    \draw (-1,0) -- (1,0)
    (-0.5,0) -- (-0.5,-1)
    (0.5,0) -- (0.5,-1)
    ;
    \filldraw[fill=Lavender] (0.3, -0.25) rectangle node {$\sigma_r$} (0.7, -0.65);
    \filldraw[fill=Lavender] (-0.3, -0.25) rectangle node {$\sigma_l$} (-0.7, -0.65);
    \draw node at (-0.5,0.3) {$A$};
    \draw node at (0.5,0.3) {$B$};
    \filldraw[fill=LimeGreen] (0,0) circle (0.25cm) node{$\Lambda_0$};
    \draw (-1,-1) -- (1,-1)
    (-1,-1) -- (-1,0)
    (1,-1) -- (1,0)
    ;
    \draw node at (-0.5,-1.35) {$\bar A$};
    \draw node at (0.5,-1.35) {$\bar B$};
    \filldraw[fill=LimeGreen] (0,-1) circle (0.3cm) node{$\Lambda_0^3$};
    \filldraw[fill=SkyBlue!50!LimeGreen] (-0.625,-1.125) rectangle (-0.375, -.875);
    \filldraw[fill=SkyBlue!50!LimeGreen] (-0.625,-.125) rectangle (-0.375, .125);
    \filldraw[fill=SkyBlue!50!LimeGreen] (0.375,-.125) rectangle (0.625, .125);
    \filldraw[fill=SkyBlue!50!LimeGreen] (0.375,-1.125) rectangle (0.625, -.875);
\end{tikzpicture}
}

\newcommand{\gradUV}{
\begin{tikzpicture}
    \draw (-1,0) -- (1,0)
    (-0.5, 0.25) -- (-0.5,-1.25)
    (0.5, 0.25) -- (0.5,-1.25)
    ;
    \filldraw[fill=Lavender] (0.3, -0.25) rectangle node {$\sigma_r$} (0.7, -0.65);
    \filldraw[fill=Lavender] (-0.3, -0.25) rectangle node {$\sigma_l$} (-0.7, -0.65);
    \draw node at (-0.75,0.25) {$U$};
    \draw node at (0.75,0.25) {$V$};
    \draw node at (0.5,0.5) {\tiny $\ket{0}$};
    \draw node at (-0.5,0.5) {\tiny $\ket{0}$};
    \filldraw[fill=gray] (0.6, 0.25) arc (0:180:0.1)
    (0.4, 0.25) -- (0.6, 0.25)
    ;
    \filldraw[fill=gray] (-0.4, 0.25) arc (0:180:0.1)
    (-0.4, 0.25) -- (-0.6, 0.25)
    ;
    \filldraw[fill=LimeGreen] (0,0) circle (0.25cm) node{$\Lambda_0$};
    \draw (-1,-1) -- (1,-1)
    (-1,-1) -- (-1,0)
    (1,-1) -- (1,0)
    ;
    \draw node at (-0.75,-1.25) {$\bar U$};
    \draw node at (0.75,-1.25) {$\bar V$};
    \draw node at (0.5,-1.5) {\tiny $\ket{0}$};
    \draw node at (-0.5,-1.5) {\tiny $\ket{0}$};
    \filldraw[fill=gray] (0.6, -1.25) arc (0:-180:0.1)
    (0.4, -1.25) -- (0.6, -1.25)
    ;
    \filldraw[fill=gray] (-0.4, -1.25) arc (0:-180:0.1)
    (-0.4, -1.25) -- (-0.6, -1.25)
    ;
    \filldraw[fill=LimeGreen] (0,-1) circle (0.3cm) node{$\Lambda_0^3$};
    \filldraw[fill=NavyBlue!50] (-0.625,-1.125) rectangle (-0.375, -.875);
    \filldraw[fill=NavyBlue!50] (-0.625,-.125) rectangle (-0.375, .125);
    \filldraw[fill=NavyBlue!50] (0.375,-.125) rectangle (0.625, .125);
    \filldraw[fill=NavyBlue!50] (0.375,-1.125) rectangle (0.625, -.875);
\end{tikzpicture}
}

\newcommand{\secondmomentstepI}{
\begin{tikzpicture}
    \draw (-0.5,0) -- (1,0)
    (0.5, 0.25) -- (0.5,-1.25)
    ;
    \filldraw[fill=Lavender] (0.3, -0.25) rectangle node {$\sigma_r$} (0.7, -0.65);
    \draw node at (0.75,0.25) {$V$};
    \draw node at (0.5,0.5) {\tiny $\ket{0}$};
    \filldraw[fill=LimeGreen] (0,0) circle (0.25cm) node{$\Lambda_0$};
    \filldraw[fill=gray] (0.6, 0.25) arc (0:180:0.1)
    (0.4, 0.25) -- (0.6, 0.25)
    ;
    \filldraw[fill=gray] (0.6, -3.25) arc (0:-180:0.1)
    (0.4, -3.25) -- (0.6, -3.25)
    ;
    \filldraw[fill=gray] (0.6, -1.75) arc (0:180:0.1)
    (0.4, -1.75) -- (0.6, -1.75)
    ;
    \filldraw[fill=gray] (0.6, -1.25) arc (0:-180:0.1)
    (0.4, -1.25) -- (0.6, -1.25)
    ;
    \draw (-0.4,-1) -- (1,-1)
    (1,-1) -- (1,0)
    ;
    \draw node at (0.75,-1.25) {$\bar V$};
    \draw node at (0.25,-1.33) {\tiny $\ket{0}$};
    \filldraw[fill=LimeGreen] (0,-1) circle (0.3cm) node{$\Lambda_0^3$};

    \draw (-0.4,-2) -- (1,-2)
    (-0.4,-2) -- (-0.4,-1)
    (0.5, -1.75) -- (0.5, -3.25)
    ;
    \filldraw[fill=Lavender] (0.3, -2.25) rectangle node {$\sigma_r$} (0.7, -2.65);
    \draw node at (0.75,-1.75) {$V$};
    \draw node at (0.25,-1.7) {\tiny $\ket{0}$};
    \filldraw[fill=LimeGreen] (0, -2) circle (0.25cm) node{$\Lambda_0$};
    \draw (-0.5,-3) -- (1,-3)
    (-0.5,-3) -- (-0.5,0)
    (1,-3) -- (1,-2)
    ;
    \draw node at (0.75,-3.25) {$\bar V$};
    \draw node at (0.5,-3.5) {\tiny $\ket{0}$};
    \filldraw[fill=LimeGreen] (0,-3) circle (0.3cm) node{$\Lambda_0^3$};
    \filldraw[fill=NavyBlue!50] (0.375,-.125) rectangle (0.625, .125);
    \filldraw[fill=NavyBlue!50] (0.375,-1.125) rectangle (0.625, -.875);
    \filldraw[fill=NavyBlue!50] (0.375,-2.125) rectangle (0.625, -1.875);
    \filldraw[fill=NavyBlue!50] (0.375,-3.125) rectangle (0.625, -2.875);
\end{tikzpicture}
}

\newcommand{\unitary}[3]{
\begin{tikzpicture}
    \draw (-0.5,-0.5) -- (-0.5,0.5);
    \filldraw[fill=NavyBlue!50] (-0.75, -0.25) rectangle node[black] {#1} (-0.25, 0.25);
    \draw node at (-0.3,0.5) {#2};
    \draw node at (-0.3,-0.5) {#3};
\end{tikzpicture}
}

\newcommand{\deltaTensor}[4]{
\begin{tikzpicture}
    \draw (-0.5,-0.2) -- (0.5,-0.2)
    (-0.5,0.2) -- (0.5,0.2)
    ;
    \draw node at (-0.7,0.2) {#1};
    \draw node at (-0.7,-0.2) {#2};
    \draw node at (0.7,0.2) {#3};
    \draw node at (0.7,-0.2) {#4};
\end{tikzpicture}
}

\newcommand{\deltaTensorII}[8]{
\begin{tikzpicture}
    \draw (-0.5,-0.5) -- (0.5,-0.5)
    (-0.5,0.5) -- (0.5,0.5)
    (-0.5,-0.25) -- (0.5,-0.25)
    (-0.5,0.25) -- (0.5,0.25)
    ;
    \draw node at (-0.7,0.25) {#3};
    \draw node at (-0.7,-0.25) {#2};
    \draw node at (0.7,0.25) {#7};
    \draw node at (0.7,-0.25) {#6};
    \draw node at (-0.7,0.5) {#1};
    \draw node at (-0.7,-0.5) {#4};
    \draw node at (0.7,0.5) {#5};
    \draw node at (0.7,-0.5) {#8};
\end{tikzpicture}
}

\newcommand{\deltaTensorIII}[8]{
\begin{tikzpicture}
    \draw (-0.5,-0.5) -- (0.5,-0.25)
    (-0.5,0.5) -- (0.5,0.25)
    (-0.5,-0.25) -- (0.5,-0.5)
    (-0.5,0.25) -- (0.5,0.5)
    ;
    \draw node at (-0.7,0.25) {#3};
    \draw node at (-0.7,-0.25) {#2};
    \draw node at (0.7,0.25) {#7};
    \draw node at (0.7,-0.25) {#6};
    \draw node at (-0.7,0.5) {#1};
    \draw node at (-0.7,-0.5) {#4};
    \draw node at (0.7,0.5) {#5};
    \draw node at (0.7,-0.5) {#8};
\end{tikzpicture}
}

\newcommand{\deltaTensorIV}[8]{
\begin{tikzpicture}
    \draw (-0.5,-0.5) -- (0.5,-0.25)
    (-0.5,0.5) -- (0.5,0.5)
    (-0.5,-0.25) -- (0.5,-0.5)
    (-0.5,0.25) -- (0.5,0.25)
    ;
    \draw node at (-0.7,0.25) {#3};
    \draw node at (-0.7,-0.25) {#2};
    \draw node at (0.7,0.25) {#7};
    \draw node at (0.7,-0.25) {#6};
    \draw node at (-0.7,0.5) {#1};
    \draw node at (-0.7,-0.5) {#4};
    \draw node at (0.7,0.5) {#5};
    \draw node at (0.7,-0.5) {#8};
\end{tikzpicture}
}

\newcommand{\deltaTensorV}[8]{
\begin{tikzpicture}
    \draw (-0.5,-0.5) -- (0.5,-0.5)
    (-0.5,0.5) -- (0.5,0.25)
    (-0.5,-0.25) -- (0.5,-0.25)
    (-0.5,0.25) -- (0.5,0.5)
    ;
    \draw node at (-0.7,0.25) {#3};
    \draw node at (-0.7,-0.25) {#2};
    \draw node at (0.7,0.25) {#7};
    \draw node at (0.7,-0.25) {#6};
    \draw node at (-0.7,0.5) {#1};
    \draw node at (-0.7,-0.5) {#4};
    \draw node at (0.7,0.5) {#5};
    \draw node at (0.7,-0.5) {#8};
\end{tikzpicture}
}

\newcommand{\MPS}{
\begin{tikzpicture}
    \draw (-4,0) -- (1,0)
    (-1,0) -- (-1,0.5)
    (1,0) -- (1,0.5)
    (-2,0) -- (-2,0.5)
    (-3,0) -- (-3,0.5)
    (0,0) -- (0,0.5)
    (-4,0) -- (-4,0.5)
    ;
    \draw node at (-3.5,-0.125) {\tiny $D_1$};
    \draw node at (-2.5,-0.125) {\tiny $D_2$};
    \draw node at (-4,-0.375) {$A_{[1]}$};
    \draw node at (-3,-0.375) {$A_{[2]}$};
    \draw node at (-1,-0.375) {$...$};
    \draw node at (1,-0.375) {$A_{[n]}$};
    \filldraw[fill=SkyBlue!50!LimeGreen] (-4.125,-.125) rectangle (-3.875, .125);
    \filldraw[fill=SkyBlue!50!LimeGreen] (-3.125,-.125) rectangle (-2.875, .125);
    \filldraw[fill=SkyBlue!50!LimeGreen] (-2.125,-.125) rectangle (-1.875, .125);
    \filldraw[fill=SkyBlue!50!LimeGreen] (-1.125,-.125) rectangle (-.875, .125);
    \filldraw[fill=SkyBlue!50!LimeGreen] (-.125,-.125) rectangle (.125, .125);
    \filldraw[fill=SkyBlue!50!LimeGreen] (1.125,-.125) rectangle (.875, .125);
\end{tikzpicture}
}

\begin{document}
\title{Preparation Circuits for Matrix Product States \\ by Classical Variational Disentanglement}

\author{Refik Mansuroglu \orcidlink{0000-0001-7352-513X}}
\email[]{Refik.Mansuroglu@univie.ac.at}
\affiliation{University of Vienna, Faculty of Physics, Boltzmanngasse 5, 1090 Wien, Austria}

\author{Norbert Schuch \orcidlink{0000-0001-6494-8616}}
\affiliation{University of Vienna, Faculty of Physics, Boltzmanngasse 5, 1090 Wien, Austria}
\affiliation{University of Vienna, Faculty of Mathematics, Oskar-Morgenstern-Platz 1, 1090 Vienna, Austria}

\date{\today}

\begin{abstract}
\noindent
    We study the classical compilation of quantum circuits for the preparation of matrix product states (MPS), which are quantum states of low entanglement with an efficient classical description. Our algorithm represents a near-term alternative to previous sequential approaches by reverse application of a disentangler, which can be found by minimizing bipartite entanglement measures after the application of a layer of parameterized disentangling gates. Since a successful disentangler is expected to decrease the bond dimension on average, such a layer-by-layer optimization remains classically efficient even for deep circuits. Additionally, as the Schmidt coefficients of all bonds are locally accessible through the canonical $\Gamma$-$\Lambda$ form of an MPS, the optimization algorithm can be heavily parallelized. We discuss guarantees and limitations to trainability and show numerical results for ground states of one-dimensional, local Hamiltonians as well as artificially spread out entanglement among multiple qubits using error correcting codes.
\end{abstract}

\maketitle

\begin{figure*}
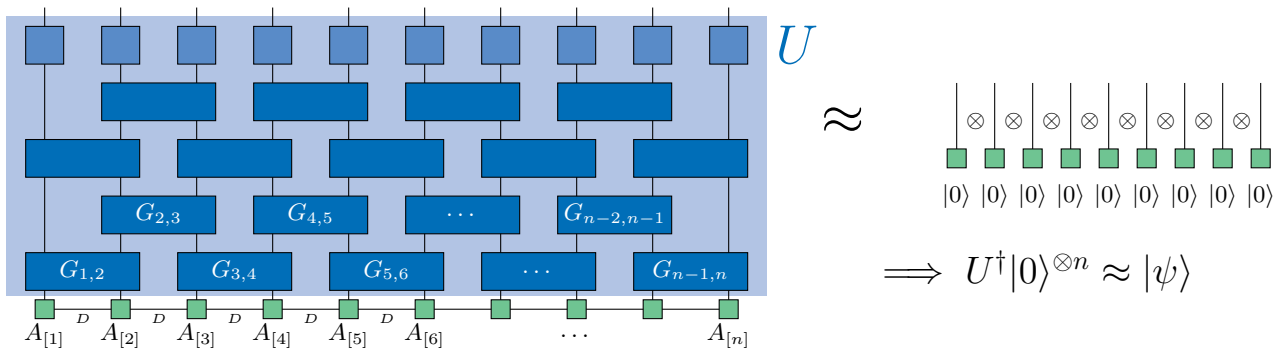

    \centering
    \disentangle
    \caption{Graphical representation of a unitary disentangler $U$ acting on an MPS $\ket{\psi}$. $U$ is built from a product of local gates $G_{i,i+1}$ (blue) acting on neighboring sites. When the bond dimension is reduced to one, a last layer of single-site gates can be read off to finally map the MPS to the computational basis state $\ket{0}^{\otimes n}$.}
    \label{fig:disentangle}
\end{figure*}

\section{Introduction}
Finding quantum circuit representations of useful entangled initial states can be decisive for a successful quantum simulation of many-body physics. But also beyond quantum simulation, mapping classical solutions of parts of the problem, or educated guesses, to quantum circuits is highly relevant in ground state problems \cite{marti2025efficientquantumcoolingalgorithm} and for avoiding barren plateaus in variational algorithms and quantum machine learning \cite{McClean_2018}.

Especially on one-dimensional chains, matrix product states (MPS) have been proven useful structures for investigations on ground states of gapped, local Hamiltonians on one-dimensional chains \cite{Verstraete_2006, Hastings_2007}, as they are efficient classical representations of quantum states with bounded bipartite entanglement. The time evolution of such systems, however, is generally a hard task \cite{Schuch_2008b}, which is believed to be a natural first candidate for quantum advantage \cite{will2025probingnonequilibriumtopologicalorder, evered2025probingtopologicalmatterfermion}. An MPS provides classical access to physical states of quantum many-body systems and is known to be exactly preparable by a quantum circuit of depth linear in system size $n$ and quadratic in the bond dimension $D$ \cite{Sch_n_2005}. More recently, it was shown that sequential preparation could be refined for MPS with finite correlation length yielding a log-depth $\mathcal{O}(D^4 \log(n))$ quantum circuit for small-error approximate preparation \cite{Malz_2024}. An interesting intermediate strategy interpolating between sequential and brickwall layouts has also been proposed \cite{wei2025statepreparationparallelsequentialcircuits} and benchmarked against other preparation strategies in the presence of noise. Using mid-circuit measurements, the depth can be further reduced to $\mathcal{O}(\log \log n)$ \cite{Malz_2024} or even constant-depth circuits \cite{Smith24}, though only for a restricted subclass of MPS, on which measurement-induced defects can be ``pushed'' to the physical indices.

While past approaches have been proven to have optimal asymptotic scaling \cite{Malz_2024}, what is needed in near-term applications are hardware-efficient circuits with 2-qubit gates on neighboring sites, allowing for approximation errors on the same scale as gate errors. In fact, a number of investigations have been built upon the sequential preparation \cite{Ran_2020, Rudolph_2024} as well as application-specific approaches like image data encoding \cite{Jobst_2024} and general classification tasks \cite{murota2025adiabaticencodingpretrainedmps}. It is important to note that while sequential preparation yields exact states, or small many-body infidelity in the log-depth method, it requires a minimum number of implementable gates to begin with. This makes it poorly suited for near-term applications, where gate budgets are severly restricted and idle qubits should be avoided, especially considering state preparation as only a small subroutine in a larger quantum algorithm. Recent approaches under the paradigm of approximate quantum compiling (AQC) have therefore proposed the general task to variationally compile circuits using tensor network methods to optimize for a target fidelity \cite{jaderberg2025variationalpreparationnormalmatrix, robertson2024approximatequantumcompilingquantum, Lubasch_2020} or to leverage counter-diabatic dynamics \cite{Mc_Keever_2024}. Other strategies include local infidelity minimization \cite{Ben_Dov_2024}, VQE pre-routines \cite{khan2023preoptimizingvariationalquantumeigensolvers}, and quantum circuit representations of analytical functions \cite{Melnikov_2023}. While these methods highlight the power of tensor-network-assisted optimization and the benefits of local cost functions for avoiding barren plateaus, fidelity-based cost function do not offer control over the bond dimension of the MPS, thus allowing a potential exponential growth of the bond dimension in circuit depth. The challenge remains to design a strategy that is both trainable and classically efficient for arbitrary circuit depths, and at the same time flexible enough to adapt to practical gate budgets on near-term devices.

In this work, we present classical variational disentanglement (CVD), a near-term focused approach to MPS preparation with parametrized brick-wall circuits optimized to disentangle the state using entanglement entropies as a cost function. At each bond, the Schmidt values can be efficiently processed to calculate von Neumann and Rényi entropies. Once a unitary disentangler $U$ as a product of local gates is found, the preparation circuit is read off as the inverse operator $U^\dagger$. While in general, contracting quantum circuits with MPS increases the bond dimension exponentially in the circuit depth, a successful disentangler decreases the bond dimension of the MPS on average and thus ensures efficient classical scaling. We further show that the approximation error by truncating an MPS, that is close to a product state, and reverse propagating it with $U^\dagger$ is bounded by the square root of the system size $\sqrt{n}$, the circuit depth $L$ and the truncation error $\epsilon$. Finally, we show that as a consequence of ultra-locality of the cost function, i.e. being defined on only two sites of the MPS, there are no barren plateaus that would impede trainability of the model. In that, CVD enhances the idea of AQC by using the local nature of the encoding of an MPS enabling efficient, local gate-by-gate optimizations.

Although previous works have explored disentangling‐based and variational approaches to prepare MPS, our work differs crucially in three respects. First, we introduce an entanglement‐based cost function, using Rényi or von Neumann entropies, whose minimization is shown to provably bound the bond dimension (Lemma \ref{lem:bond_dim}), ensuring that the algorithm remains classically efficient regardless of how deep the circuit becomes. Second, we establish trainability guarantees: together with an explicit error bound (Lemma \ref{lem:error}), we prove absence of barren plateaus (Lemma \ref{lem:barren}). Finally, our method is designed with near‐term constraints in mind—i.e. fixed gate budgets rather than exact zero‐error or asymptotic behavior—and is applicable with any circuit ansatz, and not necessarily tied to a brickwall ansatz, making it more flexible than sequential preparation or global overlap‐based optimization approaches. These features combined—bond‐dimension bounding, trainability, fixed‐budget working, and ansatz‐independence—are, to our knowledge, not present together in any previous state‐preparation strategy.

\section{Classical Variational Disentanglement}
We consider finite matrix product states (MPS) with open boundary conditions, i.e. we assume $D_0 = D_{n} = 1$:
\begin{align}
    \ket{\psi} &= \sum_{b_1, ..., b_n} A_{[1]}^{b_1} \cdots A_{[n]}^{b_n} \ket{b_1 \dots b_n} \nonumber \\
    &= \vcenter{\hbox{\MPS}}.
    \label{eq:MPS}
\end{align}
The matrices $A_{[k]} \in \Mat( D_{k-1} \times D_k, \mathds{C})$ have a maximal bond dimension $D = \max_k D_k$, which is assumed to be constant in system size. As most of the discussion is considering local quantities, CVD can also be applied to map finite chunks of infinitely large systems to quantum circuits. In this case, we can allow for larger bond dimensions at the boundary and contract $A_{[1]}$ and $A_{[n]}$ with $D_0$ and $D_n$ dimensional vectors $\ket{\phi_{l/r}}$ for the left and right boundary, respectively. On systems with periodic boundary conditions, the concepts can be similarly applied, although the calculation of the entanglement entropies is more tedious, as there does not exist a canonical form, in general.

\subsection{Overview}
In order to minimize entanglement entropies on all bonds simultaneously, we are working with Vidal's $\Gamma$-$\Lambda$ form \cite{Vidal_2003, Schollw_ck_2011}, in which the $\Lambda$ matrices are diagonal, positive definite matrices carrying the Schmidt values of the bipartite state separated by the respective bond. A finite MPS $\ket{\psi} \in \mathcal{H}$ can always be brought into $\Gamma$-$\Lambda$ form,
\begin{align}
    \ket{\psi} &= \sum_{b_1, ..., b_n} \Gamma_1^{b_1} \Lambda_1 \Gamma_2^{b_2} \Lambda_2 \cdots \Lambda_{n-1} \Gamma_n^{b_n} \ket{b_1 \dots b_n} \nonumber
\end{align}
\begin{align}
    &= \vcenter{\hbox{\MPSGL}},
\end{align}
by applying a singular value decomposition (SVD) on each site \cite{Schollw_ck_2011}. We are looking for a quantum circuit $U = \prod_i G_{i,i+1}$ built from a product of gates $G_{i,i+1} \in $ SU($d^2$), that maps a given MPS $\ket{\psi} \in \mathcal{H} \cong (\mathds{C}^d)^{\otimes n}$ as above to a product state. For the sake of concreteness, we focus on the qubit case, $d=2$, but CVD works the same way on systems with higher local dimension $d$ if the gates $G_i \in {\rm SU}(d^2)$ are adapted adequately.

Once a product state is reached, it can be mapped to $\ket{0}^{\otimes n}$ by one layer of single-qubit gates (see Fig.~\ref{fig:disentangle}). We will absorb this last layer into $U$ without loss of generality. This is possible since a product state, i.e. an MPS with bond dimension $D=1$ can be prepared with a Pauli X and Pauli Z rotation with angles $\phi_{x/z}$ that can be read off of the coefficients $\Gamma^0 = \cos(\phi_x)$ and $\Gamma^1 = e^{i\phi_z}\sin(\phi_x)$. The disentangler $U$ automatically leaves us with a state preparation circuit simply by reverse application.

To ensure classical efficiency, the bond dimensions of the MPS is truncated under a map $\mathcal{T}_D: \mathcal{H} \to \mathcal{H}$ that projects the input state to the closest bond dimension $D$ MPS by cutting off all but the largest $D$ Schmidt coefficients. Even if the disentangler does not create a perfect product state after $L$ layers, we can terminate CVD by applying the truncation map $\mathcal{T}_1$ and reading off the last layer of single-qubit gates as described above. The state that would be ultimately prepared on a quantum computer, $U^\dagger \ket{0}$, carries an error that is discussed in detail in Lemma ~\ref{lem:error}, where we prove an upper bound containing the truncation error
\begin{align}
    \epsilon &= \norm{\ket{\psi} - \left( \prod_{i=L}^1 \mathcal{T}_D \circ G_i^\dagger \right) \circ \mathcal{T}_1 \circ \left( \prod_{i=1}^L \mathcal{T}_D \circ G_i \right) \ket{\psi}} \nonumber \\
    &= \norm{\ket{\psi} - \left( \prod_{i=L}^1 \mathcal{T}_D \circ G_i^\dagger \right) \ket{0}},
    \label{eq:overlap_error}
\end{align}
which is calculated by forward application, truncation to the product state $\ket{0}$ (omitting the last layer of single-qubit gates without loss of generality) and finally reverse application of the disentangler while keeping a maximum bond dimension $D$. We relabeled the gates $G_i$ to be ordered in the brick-wall structure as depicted in Fig.~\ref{fig:disentangle}. The product $\prod$ orders its terms from right to left, i.e. $\prod_{i=1}^L G_i$ applies $G_1$ first and $G_L$ last, and $\prod_{i=L}^1$ is reverse. This error is classically accessible and can thus be used as an additional figure of merit.

\subsection{Disentanglement by Entropy Minimization}
In $\Gamma$-$\Lambda$ form, an efficiently calculable entanglement measure on $\ket{\psi}$ is the $\alpha$-Rényi entanglement entropy, $S_{\mathcal{A}, \alpha} (\theta) = \frac{1}{1-\alpha} \log\left( \Tr(\rho_\mathcal{A}^\alpha) \right)$, of a bipartition into a subsystem $\mathcal{A}$ and its complement $\mathcal{A}^c$. A natural choice for $\mathcal{A}$ and $\mathcal{A}^c$ on an MPS is defined by the $i^\text{th}$ bond, whose $\Lambda$ matrix contains the Schmidt values of a Schmidt decomposition with respect to the bipartition into $[i] := \{j | j \leq i\}$ and its complement. The Rényi entropy can thus be calculated without explicitly working with the reduced density matrix $\rho_{\mathcal{A}}$, it reads
\begin{align}
    S_{\mathcal{A}, \alpha} &= \frac{1}{1-\alpha} \log\left( \Tr(\rho_\mathcal{A}^\alpha) \right) \nonumber \\
    &= \frac{1}{1-\alpha} \log\left( \sum_j \lambda_j^{2\alpha} \right),
    \label{eq:renyi_entropy}
\end{align}
with $\lambda_j := \Lambda_{jj}$ being the $j^\text{th}$ diagonal element of the $\Lambda$ matrix defining the bipartition. In order to find the disentangler, we are treating $S_{\mathcal{A}, \alpha}$ as a cost function for variational optimization on the variational reduced density matrix $\rho_\mathcal{A}(\theta) = \Tr_{\mathcal{A}^c}(U(\theta) \ket{\psi} \bra{\psi} U(\theta)^\dagger)$, or in other words on the variational $\Lambda(\theta)$ matrices. As an ansatz for $U(\theta)$, we choose a brick wall circuit, 
\begin{align}
    U(\theta) = \prod_{\mu = 1}^L \left( \prod_{i \, \rm{odd}}^n G^\mu_{i, i+1} (\theta) \prod_{i \, \rm{even}}^n G^\mu_{i, i+1} (\theta) \right),
\end{align}
consisting of $L$ layers of the most general disentangler \cite{Kraus_2001} on two qubits, that is
\begin{align}
    &G_{i,j}(\theta) = e^{-i (\theta_{i,1} X_iX_j + \theta_{i,2} Y_iY_j + \theta_{i,3} Z_i Z_j)} \times \nonumber \\
    &\times \left( e^{-i (\theta_{i,4} X_j + \theta_{i,5} Y_j + \theta_{i,6} Z_j)} \otimes e^{-i (\theta_{i,7} X_i + \theta_{i,8} Y_i + \theta_{i,9} Z_i)} \right).
\end{align}
The most general SU(4) operator from \cite{Kraus_2001} also includes single-qubit gates at the end. These do not need to be part of $G_{i,i+1}$, as they do not remove any entanglement. With 9 parameters per gate, the total number of parameters is hence $9(n-1)L$. Although this is the most general ansatz, in some applications it might be favorable to choose a more restricted gate set tailored to the available hardware, accepting possibly deeper circuits. In that case, the method can be straightforwardly adapted. 

To ensure a classically efficient evaluation of the cost function, we perform a layer-by-layer optimization of the parameters $\theta$, since a contraction of the full circuit on non-optimized parameters can lead to an exponential scaling of the bond dimension $D \lesssim 4^L$. The contraction of a single disentangling gate with two sites of the input MPS, on the other hand, is efficient and can be parallelized for every gate in the same layer. After contracting a gate $G_{l,r}$ on two sites $l$ and $r$, the new Schmidt values $\Lambda_0'$ result from an SVD in the following way:
\begin{align}
    &\vcenter{\hbox{\resizebox{4cm}{1.6cm}{\gateActionI}}} = \vcenter{\hbox{\resizebox{4cm}{1.6cm}{\gateActionII}}} \nonumber \\
    &= 
    \vcenter{\hbox{\resizebox{3cm}{0.75cm}{\gateActionIV}}} =: \vcenter{\hbox{\resizebox{4cm}{1.4cm}{\gateActionV}}}.
    \label{eq:apply_gate}
\end{align}
Using a matrix product operator (MPO) form of the gate $G_{l,r}$ shows that the contraction of $\Gamma_{l/r}$ with $G_{l,r}$ can be carried out efficiently paying the prize of an increase in bond dimension of $\Lambda_0$ by a factor of 4 in the worst case. In the second step, all tensors are being contracted and isometries $U_l$ and $V_r$ as well as the new singular values $\Lambda_0'$ have been defined via singular value decomposition (SVD). Finally, the new $\Gamma$ tensors are defined via
\begin{align}
    \Gamma_l' = \Lambda_l^{-1} U_l \qquad \text{and} \qquad \Gamma_r' = V_r \Lambda_r^{-1}.
\end{align}
This allows for an efficient evaluation of the new entanglement entropy by plugging $\Lambda_0'$ into Eq.~\eqref{eq:renyi_entropy}. While the bond dimension could in principle grow by a factor of 4 in each such layer, the optimizer actively tries to reduce the rank of $\Lambda_0$ and therefore also the necessary bond dimension by minimizing the entanglement entropy. We give a rigorous explanation of this in Lemma \ref{lem:bond_dim}. 

\begin{figure*}
    \centering \textbf{Bond Dimensions -- Ising Model}
    \begin{tabular*}{\textwidth}{c@{\extracolsep{\fill}}cc}
        \hspace{2cm} $\alpha=0.5$ & $\alpha = 1$ & $\alpha = 2$ \hspace{2cm} \\
        \hspace{2cm} $\epsilon = 1.087 \times 10^{-4}$ & $\epsilon = 2.306 \times 10^{-4}$ & $\epsilon = 3.253 \times 10^{-4}$ \hspace{2cm} \\
    \end{tabular*}
    \includegraphics[width=\linewidth]{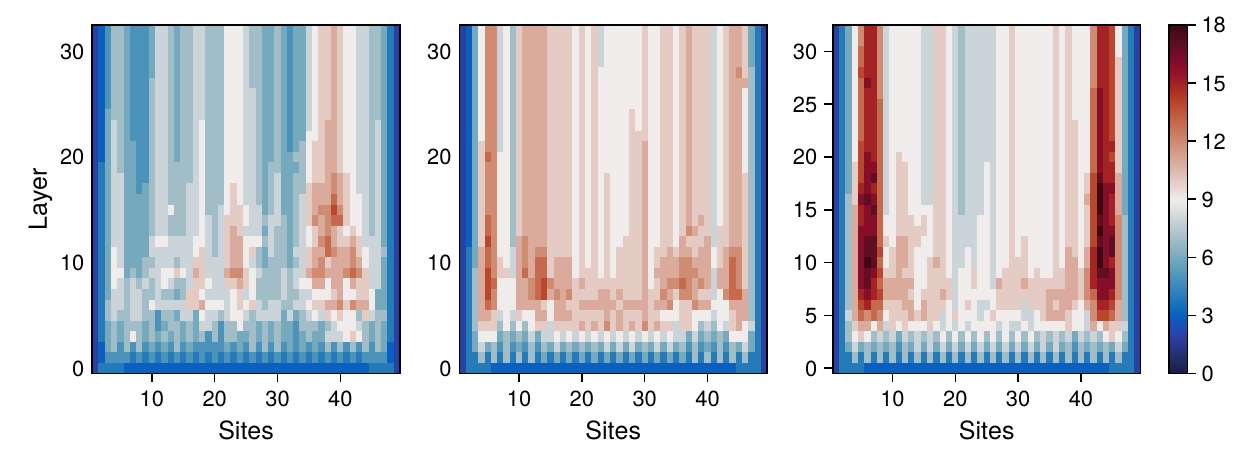}
    \caption{Bond dimensions for disentanglement of the ground state of a 1D Ising Model with skew magnetic field of strengths $h_x = 0.5$ and $h_z = 0.05$, see Eq.~\eqref{eq:TFIM}. After the target MPS is prepared with a DMRG, CVD is carried out using different values for the Rényi index $\alpha$ showcasing a suppression of the maximal bond dimension $D$ for small indices $\alpha$ in the presence of an error threshold $p = 10^{-7}$. 
    A bond dimension larger than one should not be interpreted as a failure of disentangling. All three algorithms converge to a per-site overlap error of approximately $\epsilon = 10^{-4}$. }
    \label{fig:alphas}
\end{figure*}

It is clear that the Rényi index has an influence on the change in bond dimension. In the limit $\alpha \to 0$, $S_{\mathcal{A}, \alpha}$ reduces to $\log(D)$, whereas for $\alpha \to \infty$, the entropy only considers the largest Schmidt value $S_{\mathcal{A}, \alpha} \to -\log(\lambda_1^2)$. The Rényi index thus serves as an adaptive control parameter to balance the exploration of entangled states and classical efficiency via a maximal bond dimension. Fig.~\ref{fig:alphas} illustrates the application of CVD to the ground state of a transverse field Ising model on a one-dimensional chain using different Rényi indices. While the precision of state preparation does not differ significantly, the suppression of the bond dimension by a small Rényi index becomes apparent. 
Note that a bond dimension larger than one does not imply a failure of the disentangling procedure or residual entanglement. While reducing the bond dimension ensures classical representability, the relevant figure of merit is the fidelity, which is high in all cases of Fig.~\ref{fig:alphas}. We refer to section \ref{sec:numerics} for a more detailed discussion on numerical experiments. In the following section, we show how a successful minimization of $S_{\mathcal{A}, \alpha}$ yields guarantees for classical efficiency.

\subsection{Classical Efficiency}
We prove a variant of Lemma 2 from \cite{Verstraete_2006} to give an explicit form of the bound on the bond dimension, given an entanglement entropy value as output of the optimizer.
\begin{lemma}
    \label{lem:bond_dim}
    Let $S_{\mathcal{A}, \alpha} (p_i) = \frac{1}{1-\alpha} \log\left( \sum_i p_i^{\alpha} \right)$ with $p_i := \lambda_i^2$ be the $\alpha$-Rényi entropy with $0 < \alpha <1$ given by the $\Lambda$ matrix separating an MPS into left- and right-subsystems $\mathcal{A}$ and $\mathcal{A}^c$. Let $p = \sum_{i=D+1}^{2^n} p_i$ be the error from cutting off the squared Schmidt coefficients at bond dimension $D$. Then,  
    \begin{align}
        D \leq \frac{(1-\alpha) \alpha^{\frac{\alpha}{1-\alpha}}}{p^{\frac{\alpha}{1-\alpha}}} e^{S_{\mathcal{A}, \alpha}}.
        \label{eq:max_rank}
    \end{align}
\end{lemma}
Lemma \ref{lem:bond_dim} is derived by minimization of the entanglement entropy, while considering the normalization constraint and a minimal cutoff threshold for the Schmidt values, see Appendix \ref{app:bond_dim} for details. 

We observe that for $S_{\mathcal{A}, \alpha} \to 0$, the bound on the rank goes to a constant that can be larger than one depending on $p$. Since $S_{\mathcal{A}, \alpha} = 0$ implies $D=1$, the bound becomes loose in this limit. Eq.~\eqref{eq:max_rank} supports the claim that an optimization for entropies with small $\alpha$ focus on lowering the bond dimension, since the $\alpha$-dependent pre-factor becomes small for small $\alpha$ and diverges for $\alpha \to 1$ like $\left( \frac{1}{p} \right)^{\frac{\alpha}{1 - \alpha}}$. 

In case of arbitrary precision, $p\to 0$, the rank is unbounded. To ensure efficiency, the MPS has to be truncated regularly introducing the approximation error $p$ at each bond. The unitary disentangling process $U$ thus becomes $\prod_{i=1}^L \mathcal{T}_D \circ G_i$ with the truncation map $\mathcal{T}_D$ enabling the classical calculation of an error measure $\epsilon$ from Eq.~\eqref{eq:overlap_error}. However, a state preparation on a quantum machine does not contain truncations $\mathcal{T}_D$, but prepares $U^\dagger \ket{0}$. We thus tackle the question about the total error being made in CVD when keeping a constant bond dimension $D$.
\begin{lemma}
\label{lem:error}
    Let $U = \prod_i^L G_i$ be a unitary disentangler resulting from a CVD with a precision $\epsilon = \norm{\ket{\psi} - \left( \prod_{i=L}^1 \mathcal{T}_D \circ G^\dagger_i \right) \ket{0} } $ and maximal bond dimension $D$, then the total error of state preparation is
    \begin{align}
        \norm{ \ket{\psi} - U^\dagger \ket{0} } \leq \epsilon + L\sqrt{2(n-1)p},
        \label{eq:error_bound}
    \end{align}
    with $p=\max_k p^{[k]}$ and $p^{[k]}$ denoting the truncation error at the $k^\text{th}$ bond.
\end{lemma}
See Appendix \ref{app:error} for a proof.

\subsection{Absence of Barren Plateaus}
The guarantees for classical efficiency rely on the performance of the optimizer finding transformations which lower the entanglement entropy. For a complete discussion on classical efficiency, we include a note on trainability of the variational ansatz due to the existence of significant gradients. Quantum variational algorithms are known to suffer from barren plateaus \cite{McClean_2018, cerezo2024doesprovableabsencebarren}, which ultimately leads to a sample problem in resolving the direction of the gradient. While in classical computations with MPS, the full state is available, an exponentially small gradient is not necessarily a sample problem, but still imposes an overhead in the convergence of the optimization. In the following, we show that variational disentangling does not suffer from barren plateaus, which becomes apparent from the locality of Eq.~\eqref{eq:apply_gate}, pointing out the decomposition of the problem into gate-by-gate optimizations on two sites. In order to be able to calculate expressions of the gradient, we focus on the 2-Rényi entropy and its gradient that reads
\begin{align}
    \partial_i S_{\mathcal{A}, 2} (\theta) &= - \frac{2}{\Tr(\rho_\mathcal{A}(\theta)^2)} \Tr(\rho_\mathcal{A}(\theta) \partial_i \rho_\mathcal{A}(\theta)) \nonumber \\
    \text{with } \partial_i \rho_\mathcal{A}(\theta) &= \Tr_{\mathcal{A}^c} \left( (\partial_i U(\theta)) \ket{\psi} \bra{\psi} U(\theta)^\dagger + H.c. \right). 
    \label{eq:gradient}
\end{align}
The calculation straightforwardly generalizes to Rényi indices $\alpha > 1$, however gradients might become unstable for $\alpha \leq 1$ in the presence of small Schmidt values. However, this is not a problem in practice, since the introduction of a cutoff threshold leads to the truncation of small Schmidt values before entering a singularity. At $\theta = 0$, the elements of $\partial_i \rho_\mathcal{A}(\theta)$ are of the form $-i\Tr_{\mathcal{A}^c}\left( \comm{\sigma_l \otimes \sigma_r}{\ket{\psi} \bra{\psi}} \right)$ with $\sigma_{l/r}$ having support across the boundary between $\mathcal{A}$ and $\mathcal{A}^c$. An evaluation of Eq.~\eqref{eq:gradient} in graphical notation allows a formal identification of the extrema. Next to product states ($\Lambda_0 = 1$) being minima and maximally entangled states being maxima, states with real-valued coefficients represent saddle points. This property is discussed in detail in Appendix \ref{app:barren} and ensures that a gradient calculated from finite differences will thus not halt at the unstable fixed points, but move towards product states, as desired, or possibly get stuck in local minima. 

In parameter spaces of large dimension, however, cost landscapes with small typical gradients, or barren plateaus, are a more serious concern than local minima would be. In order to make a statement about generic MPS with a fixed bond dimension $D$, we calculate the first and second moment of the gradient from Eq.~\eqref{eq:gradient} evaluated on randomly distributed MPS with bond dimension $D$. Using the isometric property of $\Lambda_l \Gamma_l$ and $\Gamma_r \Lambda_r$, we can utilize statistical tools for the Haar measure of SU(2$D$), leading to a quantification of the variance of the gradient $\partial_i S_{\mathcal{A}, 2}$ in terms of the bond dimension.
\begin{lemma}
    \label{lem:barren}
    Let $A = \Lambda_l \Gamma_l$ and $B = \Gamma_r \Lambda_r$ be random isometries with fixed bond dimension $D$ and $\Lambda_0$ a matrix of Schmidt values. The expected value of the gradient $\partial_i S_{\mathcal{A}, 2}\Big\vert_{\theta = 0}$ is zero and the variance is
    \begin{align}
        \mathds{E} \left[ \left( \partial_i S_{\mathcal{A}, 2} \Big\vert_{\theta = 0} \right)^2 \right] = \frac{8}{(2D+1)^2} \left( 1 - e^{2(S_{\mathcal{A}, 2} - S_{\mathcal{A}, 3})} \right),
    \end{align}
    for $i$ corresponding to $\sigma_l \otimes \sigma_r \in \{X\otimes X, Y \otimes Y, Z \otimes Z\}$.
\end{lemma}
The proof consists of a straightforward, but technical calculation in Weingarten calculus involving integrals over the Haar measure on SU(2$D$) to account for the random isometries $A$ and $B$. Technical details are provided in Appendix \ref{app:barren}.

The statement of Lemma \ref{lem:barren} is more general than it may first appear. The restriction to $\theta=0$ can be lifted as long as a constant maximal bond dimension is ensured by regularly truncating the state. Doing so, an error bounded by Eq.~\eqref{eq:error_bound} is introduced. Typically, barren plateaus are diagnosed by randomizing over the parameters $\theta$ rather than the state that is acted upon. One can view Lemma \ref{lem:barren} as an equivalent perspective whenever the unitary $G_{l,r}(\theta)$ is absorbed into the tensors $A$ and $B$. In that case, one would have to consider a bond dimension $4D$. While this argument ensures the absence of barren plateaus, local minima can still occur. However, in practice these are mitigated by alternating between even and odd pairs after each layer, which maintains optimization momentum and reduces the chance of getting trapped.

Furthermore, it is important to distinguish the local nature of CVD from approaches based on the quantum marginal problem \cite{Ludvik}. Starting from an MPS in canonical form allows us to locally access entanglement properties across bipartitions separating all sites to the left and right of a given bond. The corresponding local MPS tensors generally encode more information than the two-site reduced density matrix on which the optimized unitary acts.

The key distinction is that disentangling a two-site marginal addresses only strictly local correlations, whereas the MPS representation captures correlations on finite length scales beyond nearest neighbors. We refer to Sec.~\ref{sec:logical} for illustrative examples.

\begin{figure*}
    \centering
    \begin{tabular*}{\textwidth}{c@{\extracolsep{\fill}}cc}
        \textbf{(a)} \hspace{1.25cm} \textbf{Ising Model} & \textbf{XY Model} & \textbf{XXZ Model} \hspace{1.75cm} \\
        \hspace{1.8cm} \scriptsize{$S_\infty = - \log(\lambda_1^2)$} & \hspace{0cm} \scriptsize{$S_\infty = - \log(\lambda_1^2)$} & \scriptsize{$S_\infty = - \log(\lambda_1^2)$} \hspace{1.75cm} 
    \end{tabular*}
    $\vcenter{\hbox{\includegraphics[width=0.32\linewidth]{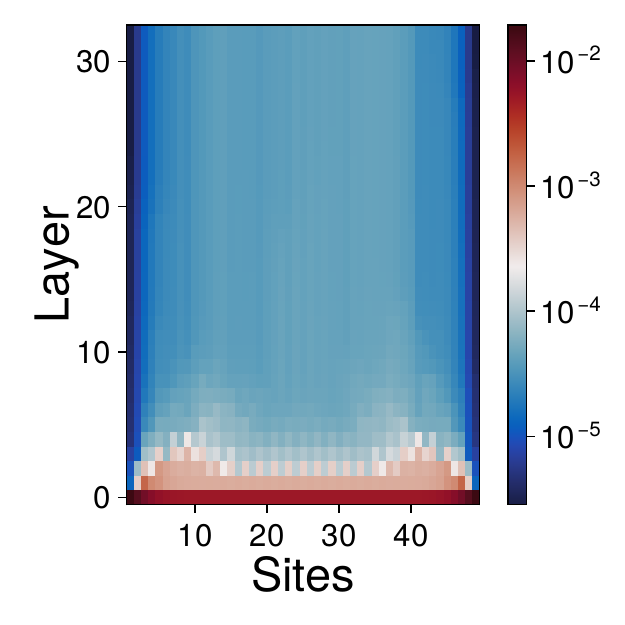}}}$
    $\vcenter{\hbox{\includegraphics[width=0.32\linewidth]{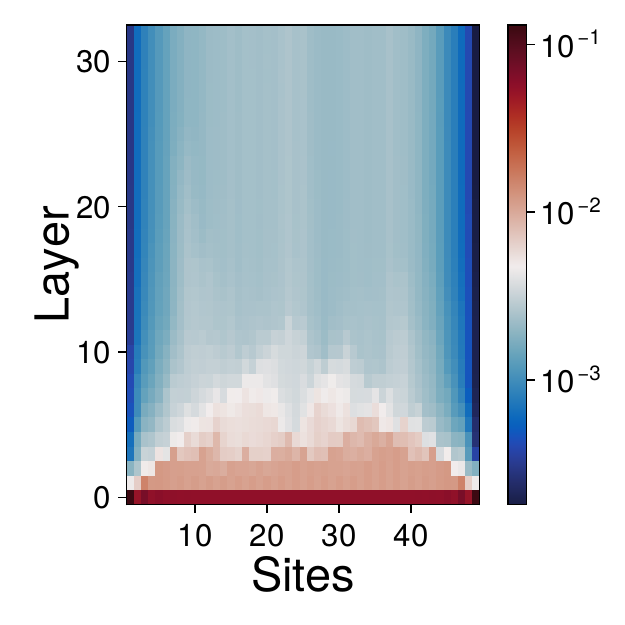}}}$
    $\vcenter{\hbox{\includegraphics[width=0.32\linewidth]{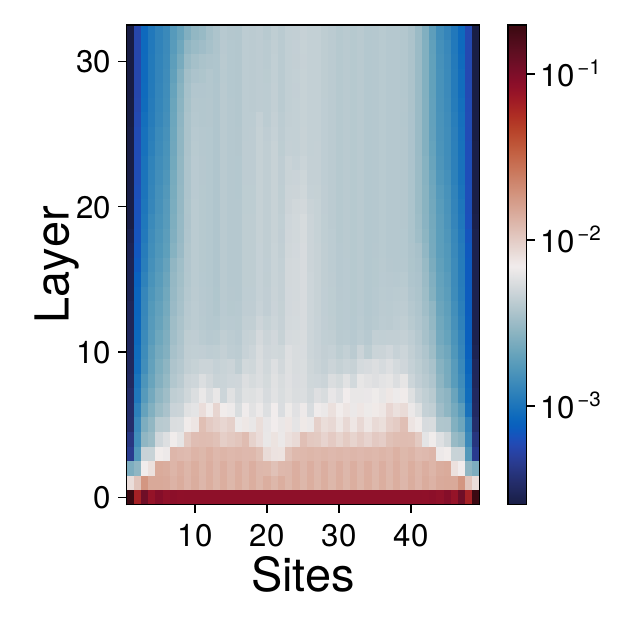}}}$
    
    \raggedright \textbf{(b)} \hspace{8cm} \textbf{(c)} 
    
    $\vcenter{\hbox{\includegraphics[width=0.45\linewidth]{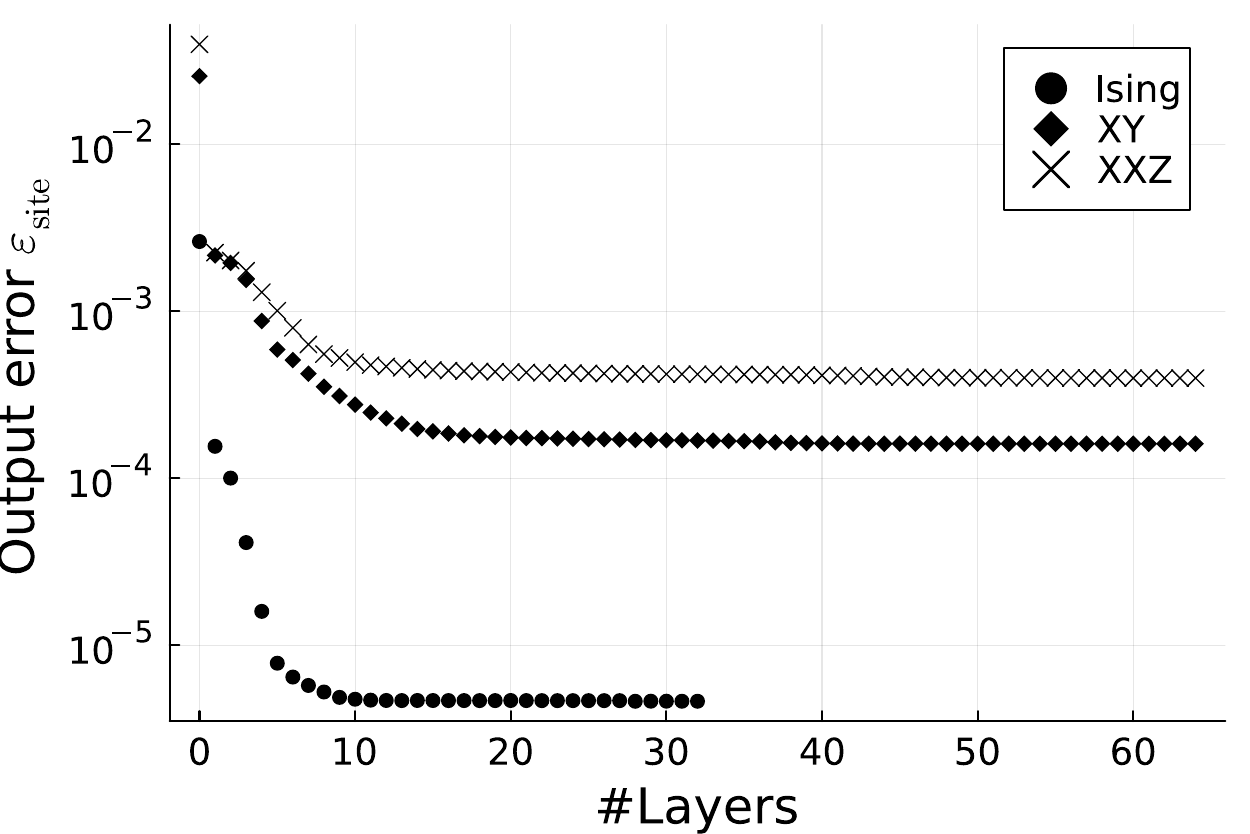}}}$
    \hspace{1cm}
    $\vcenter{\hbox{\includegraphics[width=0.45\linewidth]{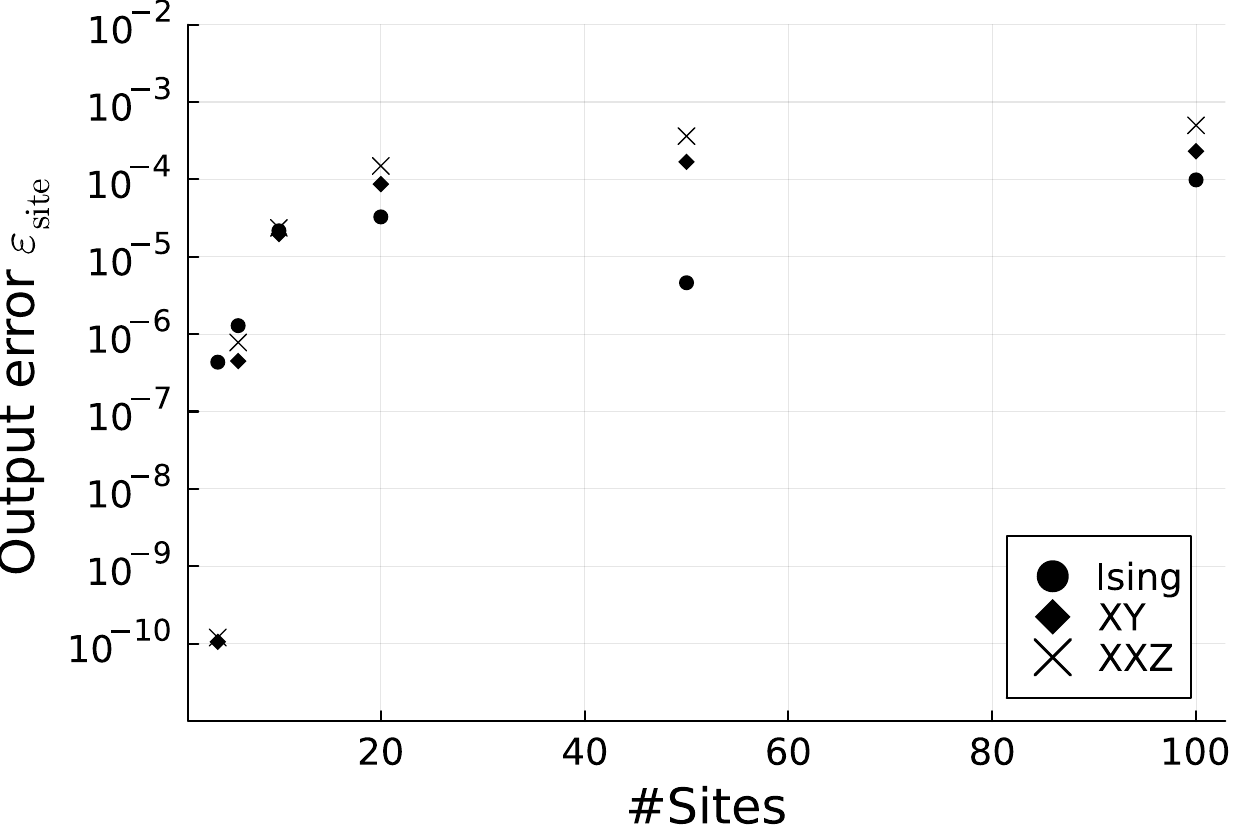}}}$
    \caption{Disentanglement of the ground states of the Ising model with skew magnetic field $h_x=0.5, h_z=0.05$, the XY Model with $g=0$ and a magnetic field in X-direction, $h_x = 0.1$, and the Heisenberg XXZ Model with $J_z = 0.5$ and a skew magnetic field $h_z = 0.1$ and $h_x = 0.1$ defined on 50 sites. The target state has been determined using a DMRG routine and subsequently disentangled using a Rényi-index $\alpha = \frac{1}{2}$ on every third layer and $\alpha = 1$ else. The color plots (a) show a gradual decrease of the $\infty$-Rényi entropy starting from the boundary. As an approximation to the tail weight, it measures the distance of the MPS to a product state across the bond. Since the overlap error (b) also saturates for all three spin models, this indicates that there exist local minima with unresolved correlations. The system size scaling (c) of the final errors after 32 layers saturate at around $10^{-4}$ with one outlier for the 50-site Ising model.}
    \label{fig:spins}
\end{figure*}

\begin{figure*}
    \centering
    \textbf{Fermi-Hubbard Model}
    
    \vspace{0.5cm}
    
    \begin{tabular*}{\textwidth}{c@{\extracolsep{\fill}}c}
        \textbf{(a)} \hspace{1.8cm} $S_\infty = - \log(\lambda_1^2)$ & \textbf{(b)} \hspace{3cm} \textbf{Energy} \hspace{3.5cm}
    \end{tabular*}
    $\vcenter{\hbox{\includegraphics[width=0.4\linewidth]{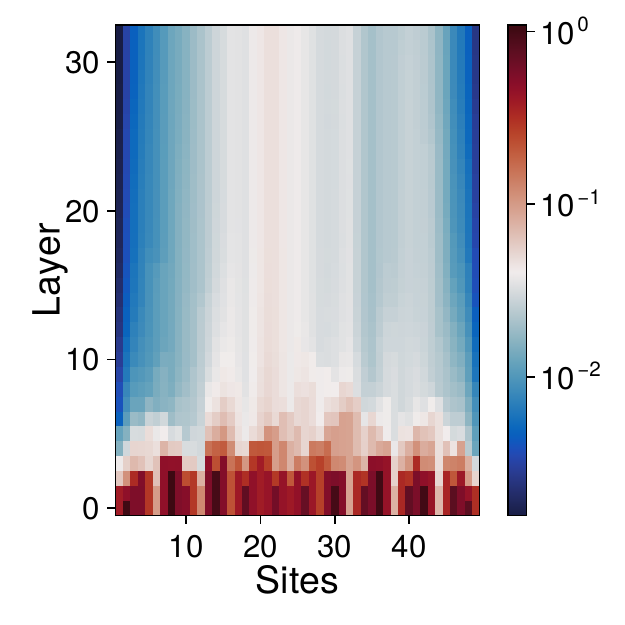}}}$
    \hspace{1cm}
    $\vcenter{\hbox{\includegraphics[width=0.5\linewidth]{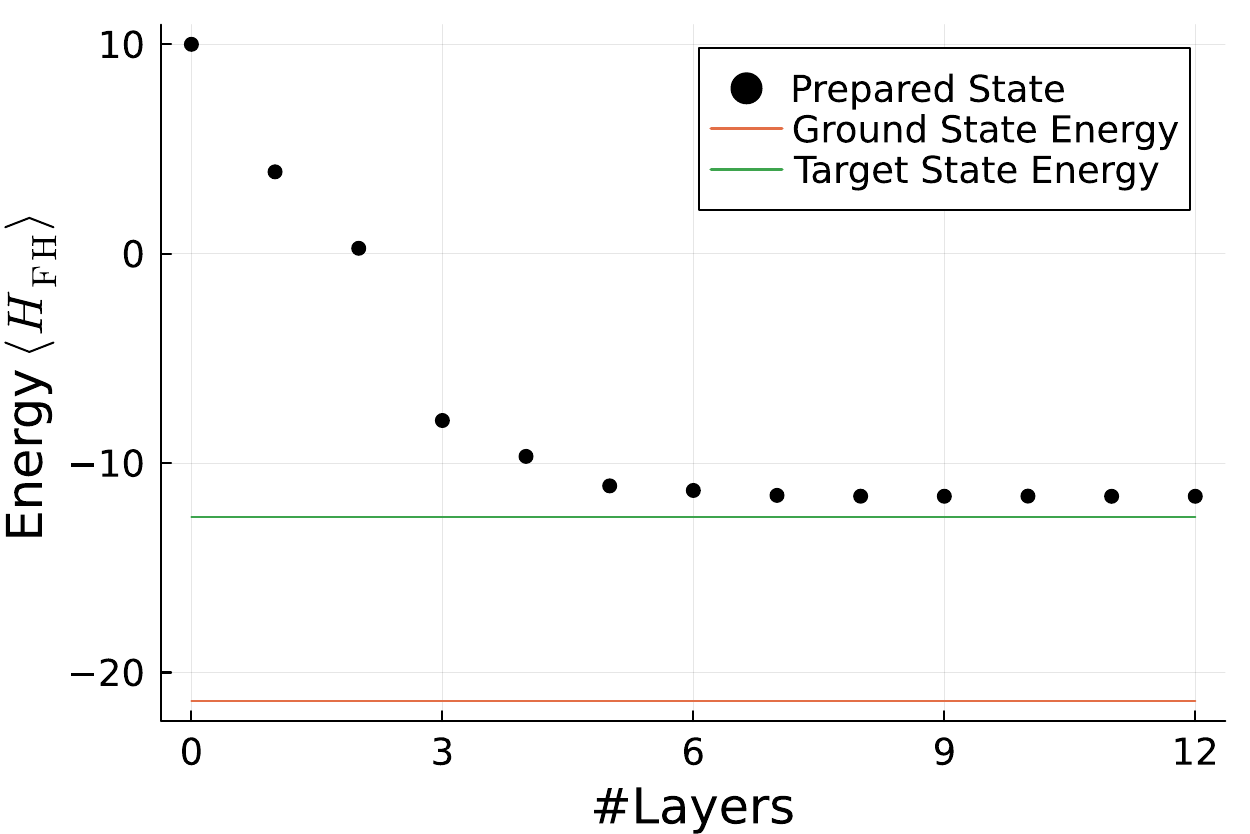}}}$
    \caption{Disentanglement of an MPS approximation to the ground state of the Fermi-Hubbard model with hopping parameter $t=1$ and Coulomb interaction strength $U=2$ on 25 electronic sites at half filling (12 spins up and 12 spins down) using $\alpha = 1$. These are encoded into 50 qubits with each qubit the occupation of a spin-up or spin-down particle, respectively. Next to a plot of the tail weight (a), the energy of the prepared low-rank state is plotted against the layer count (b). The ground state energy is approximated by a DMRG calculation with bond dimension $D=1000$ (orange) and the target state ($D=5$) energy is plotted in green. }
    \label{fig:Hubbard}
\end{figure*}

\section{Numerical Experiments}
\label{sec:numerics}
In order to test the performance and classical efficiency guarantees of CVD, we carried out a number of numerical experiments. We first consider ground states of one-dimensional spin Hamiltonians and an approximation to the ground state of a Fermi-Hubbard model found classically by the density matrix renormalization group (DMRG) method. We plot the $\infty$-Rényi entropy as a measure for the tail weight of the Schmidt values on each bond and for each layer, as well as the per-site contribution of the overlap error $\epsilon$ from Eq.~\eqref{eq:overlap_error}. The overlap error is calculated between the input MPS and an MPS resulting from forward propagation of the disentangler, truncation to bond dimension $D=1$ and finally backwards propagation of the disentangler. 

Since physical states of locally interacting spin systems are typically weakly correlated, we also test CVD on an artificial state of high bipartite entanglement. In particular, we disentangle a logical Bell pair, where two qubits are encoded in a quantum error correcting code each. Although the entanglement is distributed among many qubits, we find that CVD successfully disentangles the state only considering two neighboring sites at a time. In Appendix \ref{app:simples}, we reproduce preparation circuits for known MPS structures like the Greenberger-Horne-Zeilinger (GHZ) state, the cluster state and the Affleck–Kennedy–Lieb–Tasaki (AKLT) state exhibiting different ranges of correlation length.

All numerical experiments have been carried out with using the julia library ``ITensor'' \cite{ITensor} and can be reproduced using the code available at \cite{CVD_repo}.

\subsection{Spin Model Ground States}
An interesting quantum computing application of MPS preparation is the study of quantum many-body physics. Especially on one-dimensional spin chains, MPS have been proven faithful representations of ground states of gapped Hamiltonians with local interactions \cite{Hastings_2007, Verstraete_2006}. In this section, we find ground state preparation circuits for the following spin Hamiltonians
\begin{align}
    H_{\rm Ising} &= - \frac{1}{2} \sum_i Z_i Z_{i+1} + h_x \sum_i X_i + h_z \sum_i Z_i \label{eq:TFIM}
\end{align}
\begin{align}
    H_{\rm XY} &= - \frac{1}{2} \sum_i (X_i X_{i+1} + Y_i Y_{i+1}) + h_x \sum_i X_i 
\end{align}
\begin{align}
    H_{\rm Heisenberg} &= - \frac{1}{2} \sum_i (X_i X_{i+1} + Y_i Y_{i+1}) \nonumber \\
    &- \frac{J_z}{2} \sum_i Z_i Z_{i+1} - h_x \sum_i X_i - h_z \sum_i Z_i,
\end{align}
by first finding an MPS representation of the respective ground state using a density matrix renormalization group (DMRG) algorithm and subsequently disentangling the target MPS with CVD. There is a freedom in balancing interaction strengths in the Heisenberg model via $J_z$ and tuning the magnetic field strengths $h_x, h_z$. Choosing a state deep in a phase more likely results in extremely shallow, or even trivial preparation circuits, as the ground states can be very close to product states for some models. We thus pick near-critical models with non-trivial MPS representations and add small magnetic field values to ensure non-integrability. 

In Fig.~\ref{fig:spins}, we show results for the ground state preparation of the three spin models above. Next to the evolution of the tail weight $S_\infty = - \log(\lambda_1^2) \approx 1 - \lambda_1^2$ of the Schmidt coefficients $\lambda_i$ at any bond and after every disentangling layer, we plot the overlap error per site,
\begin{align}
    \epsilon_{\rm site} = 1 - \left( 1 - \epsilon \right)^{\frac{1}{n}},
\end{align}
with $\epsilon$ being taken from Eq.~\eqref{eq:overlap_error}. We choose this figure of merit as it reflects the error of the prepared state at the level of the individual matrices in the MPS, which contribute to the overall fidelity multiplicatively. 

The ground state of the Ising model with a skew magnetic field, $h_x=0.5$ and $h_z=0.05$, appears to have comparatively low state complexity, as it is prepared with high fidelity using around 10 brick-wall layers for a state on 50 sites. The ground states of the XY (with $h_x=0.1$) and Heisenberg XXZ models (with $J_z = 0.5$ and $h_x = h_z = 0.1$), on the other hand, converge to per-site overlap errors of just above $10^{-4}$, which is small compared to near-term gate fidelities. From a near-term perspective, it hence makes sense to stop at this point and spend the rest of the limited gate budget on the actual computation after the initial state is prepared. Higher fidelities are achievable by tuning down the cutoff parameter $p$ (compare Lemma \ref{lem:bond_dim}) and allowing for higher maximal bond dimensions along the way. The tail weights shown in Fig.~\ref{fig:spins} (a) display a gradual concentration towards a product state, as $S_\infty$ decreases starting from the boundary towards the center bonds. The maximal bond dimension of $D=100$ was never reached during CVD. 

Using the MPS representation of the ground state of a local, gapped Hamiltonian defined on a one-dimensional lattice is a very convenient position to take, as there are guarantees for a faithful representation \cite{Verstraete_2006} leaving us with practically no restriction in system size. To showcase this fact, we plot the overlap error per site, $\epsilon_{\rm site}$, of state preparations of the spin model ground states after 32 layers in Fig.~\ref{fig:spins} (c). Although the state complexities differ, resulting in different final error levels, the per-site error saturates. While it is to be expected that the total fidelity faces an orthogonality catastrophe, i.e. decreases exponentially in the system size $n$, a constant per-site error further supports the scalability of CVD. 

\begin{table*}[t]
    \raggedright \textbf{(a)} \hspace{4.2cm} \textbf{(b)}

    \centering
    \begin{tabular}{l|ccccc}
        $S_1^{[5]}$ &I &X &Z &Z &X \\
        $S_2^{[5]}$ &X &Z &Z &X &I \\
        $S_3^{[5]}$ &Z &Z &X &I &X \\
        $S_4^{[5]}$ &Z &X &I &X &Z
    \end{tabular}
    \hspace{2cm}
    \begin{tabular}{l|ccccccccccc}
        $S_1^{[11]}$ &Z&Z&Z&Z&Z&Z&I&I&I&I&I \\
        $S_2^{[11]}$ &X&X&X&X&X&X&I&I&I&I&I \\
        $S_3^{[11]}$ &I&I&I&Z&X&Y&Y&Y&Y&X&Z \\
        $S_4^{[11]}$ &I&I&I&X&Y&Z&Z&Z&Z&Y&X \\
        $S_5^{[11]}$ &Z&Y&X&I&I&I&Z&Y&X&I&I 
    \end{tabular}
    \hspace{0.5cm}
    \begin{tabular}{l|ccccccccccc}
        $S_6^{[11]}$ &X&Z&Y&I&I&I&X&Z&Y&I&I \\
        $S_7^{[11]}$ &I&I&I&Z&Y&X&X&Y&Z&I&I \\
        $S_8^{[11]}$ &I&I&I&X&Z&Y&Z&X&Y&I&I \\
        $S_9^{[11]}$ &Z&X&Y&I&I&I&Z&Z&Z&X&Y \\
        $S_{10}^{[11]}$ &Y&Z&X&I&I&I&Y&Y&Y&Z&X 
    \end{tabular}
    \caption{Stabilizers for the [5,1,3] stabilizer code \cite{laflamme1996perfectquantumerrorcorrection} (a) and the [11,1,5] stabilizer code \cite{calderbank1997quantumerrorcorrectioncodes} (b) used for the disentangling test-case shown in Fig.~\ref{fig:QEC}.}
    \label{tab:stabilizers}
\end{table*}
\begin{figure*}
    \centering
    \begin{tabular*}{\textwidth}{c@{\extracolsep{\fill}}c}
        \textbf{(a)} \hspace{3cm} \textbf{[5,1,3] code} & \textbf{(b)} \hspace{2.8cm} \textbf{[11,1,5] code} \hspace{2.8cm}
    \end{tabular*}
    \begin{tabular*}{\textwidth}{c@{\extracolsep{\fill}}ccc}
        \hspace{1cm} \textbf{separated} & \hspace{0.25cm} \textbf{interlaced} & \textbf{separated} & \textbf{interlaced} \hspace{1.25cm} \\
        \hspace{1cm} \scriptsize{$S_\infty = - \log(\lambda_1^2)$} & \hspace{0.25cm} \scriptsize{$S_\infty = - \log(\lambda_1^2)$} & \scriptsize{$S_\infty = - \log(\lambda_1^2)$} & \scriptsize{$S_\infty = - \log(\lambda_1^2)$} \hspace{1.1cm}
    \end{tabular*}
    $\vcenter{\hbox{\includegraphics[width=0.24\linewidth]{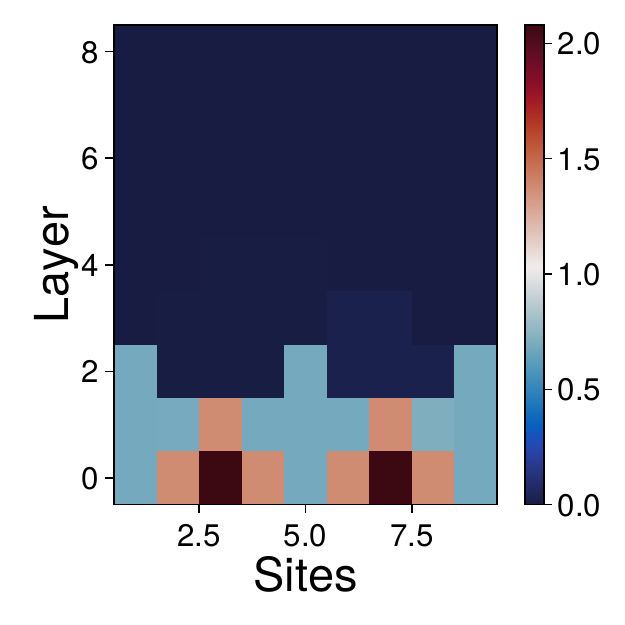}}}$
    $\vcenter{\hbox{\includegraphics[width=0.24\linewidth]{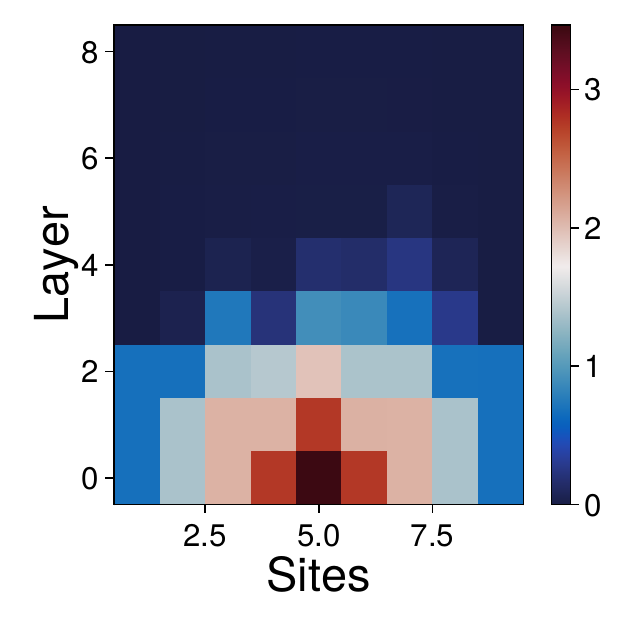}}}$
    $\vcenter{\hbox{\includegraphics[width=0.24\linewidth]{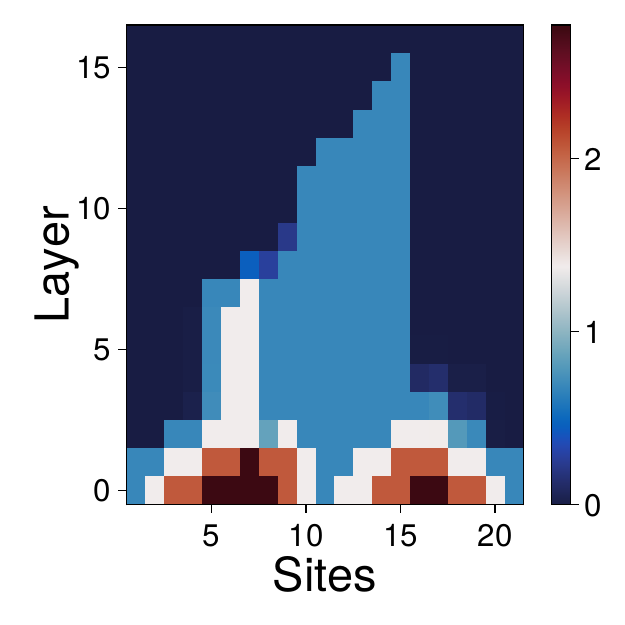}}}$
    $\vcenter{\hbox{\includegraphics[width=0.24\linewidth]{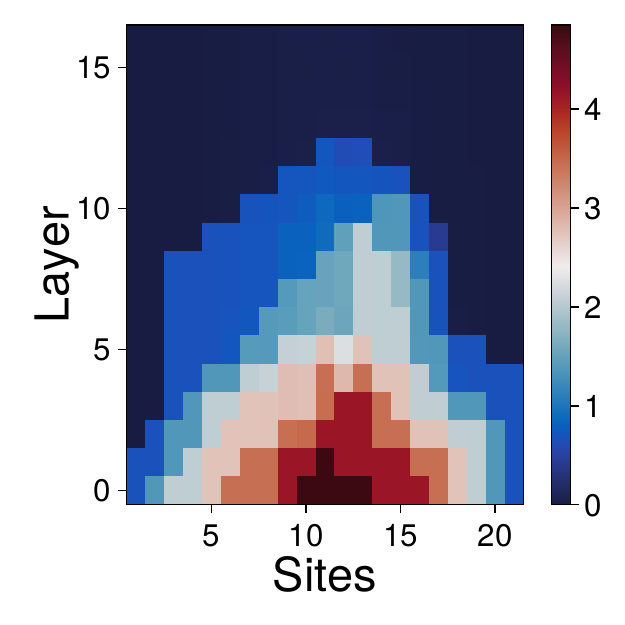}}}$
    
    \raggedright \textbf{(c)} \hspace{8cm} \textbf{(d)} 

    $\vcenter{\hbox{\includegraphics[width=0.5\linewidth]{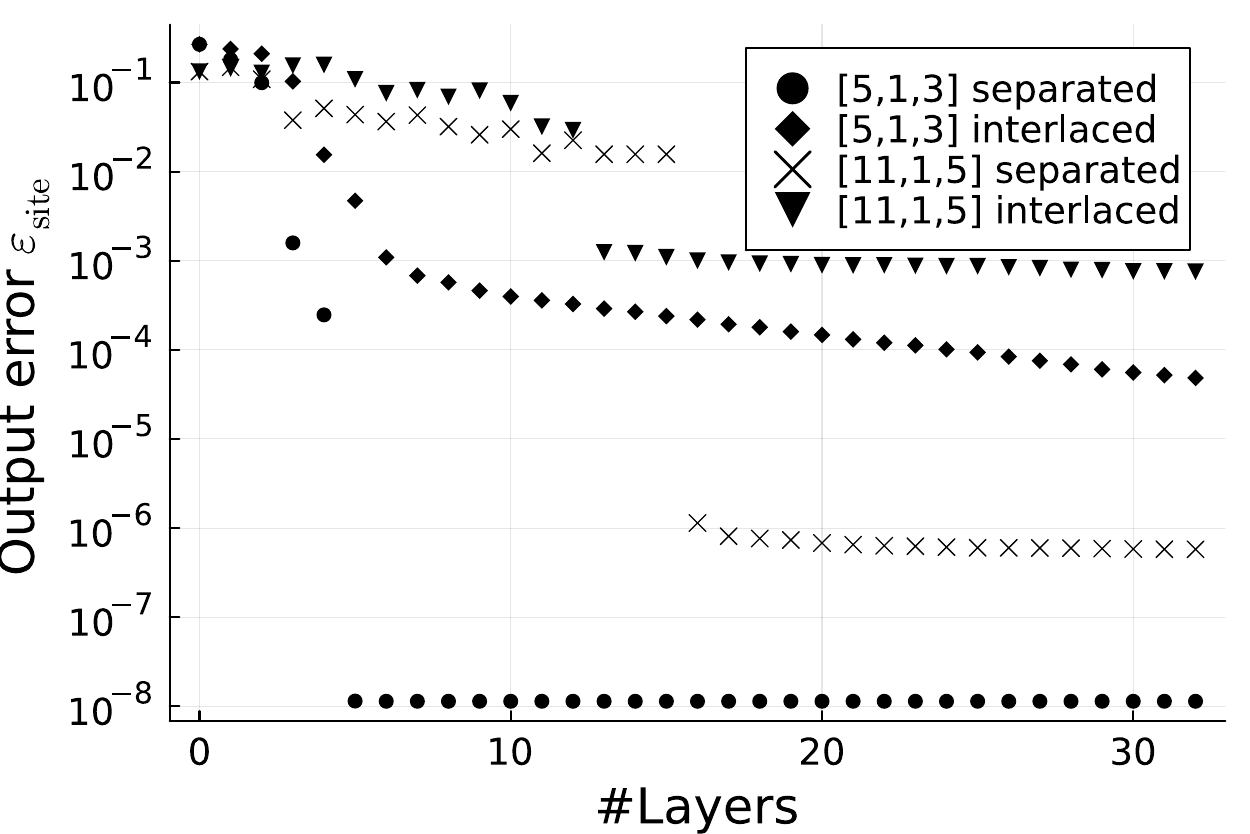}}}$ 
    $\vcenter{\hbox{\logicalbell}}$ 
    \caption{Disentanglement of logical Bell pair in a [5,1,3] and [11,1,5] stabilizer code using $\alpha = 1$ in both the separated and the interlaced version as depicted in (d). In (a) and (b), the tail weights on each bond and for each layer in the disentangling process are shown and in (c), the overlap errors from backwards preparation of the initial MPS in the [11,1,5] code are shown. We find that in all cases a linear circuit depth is required to fully disentangle the state. Although the bond dimension is decreasing along the way, the overlap error undergoes an abrupt jump.}
    \label{fig:QEC}
\end{figure*}

\subsection{1D Fermi-Hubbard Ground State}
The problem of preparing low temperature equilibrium states in the canonical ensemble of electronic structure Hamiltonians, even on one dimensional lattices, targets highly entangled states. Tensor network approaches like DMRG are struggling to find scalable approximations, in general \cite{Schuch_2009, White_2019}. This is why quantum algorithms are being investigated for the quantum simulation of electronic structure problems, but (like most quantum algorithms) they suffer from scalability problems in sample \cite{McClean_2018} and gate complexity \cite{Gonz_lez_Garc_a_2022}. An educated initial guess with sufficient overlap with the target state eases the ground state problem possibly even by an exponential factor \cite{Gharibian_2023}. This is where CVD comes into play. A (potentially faulty) DMRG and subsequent CVD yield a preparation circuit that produces a state in the low energy subspace, for which efficient ground state algorithms have been investigated, for instance via digital cooling \cite{marti2025efficientquantumcoolingalgorithm} or with variational algorithms \cite{Cade2020}. In this section, we illustrate this classical pre-processing step on the Fermi-Hubbard model with Hamiltonian
\begin{align}
    H_{\rm FH} = &-t \sum_{i, \sigma} (c^{\dagger}_{i,\sigma} c_{i+1, \sigma} + c_{i, \sigma} c_{i+1, \sigma}^{\dagger}) \nonumber \\
    &+ U \sum_i n_{i, \uparrow} n_{i, \downarrow},
\end{align}
constructed from creation and annihilation operators $c_{i, \sigma}^{\dagger}$ and $c_{i, \sigma}$ with site indices $i \in \{1, ..., n\}$ and spin indices $\sigma \in \{\uparrow, \downarrow\}$. The operators $n_{i, \sigma} := c_{i, \sigma}^{\dagger} c_{i, \sigma}$ are the number operators on site $i$ and spin $\sigma$. Running a DMRG within a particle number conserving subspace yields an MPS with local dimension 4. This state can be mapped to qubits yielding an MPS with local dimension 2, which is suitable for CVD. Each electronic site is separated into two neighboring qubit sites representing the spin-up and spin-down degree of freedom, respectively. Although the DMRG with heavily restricted bond dimension might not be quite successful in finding the ground state of $H_{\rm FH}$, it provides a low energy state as an educated guess for a potential quantum computing application. For illustration, we probe the energy of an MPS approximation with bond dimension $D=5$ of the Fermi-Hubbard groundstate on 25 electronic sites, or 50 qubit sites respectively. We compare the energy of the state prepared by local unitary gates using CVD to a benchmark of the ground state energy using DMRG on an MPS with bond dimension $D=1000$ as well as the energy of the target state ($D=5$) in Fig.~\ref{fig:Hubbard}. Even though the prepared quantum state lies far above the actual ground state in energy, it still exemplifies how CVD can generate structured initial states that may reduce the complexity of a subsequent quantum solution to the ground state problem. Importantly, our example is meant as a preparatory step rather than a proposal for solving ground state problems in 1D, where DMRG is already highly efficient. The task becomes much harder and more interesting in 2D, where the Fermi–Hubbard model is \QMA-hard \cite{Schuch_2009}, but can be reduced to \BQP \ when a groundstate guess with inverse-polynomial precision is available \cite{Gharibian_2023}. Cooling algorithms such as \cite{marti2025efficientquantumcoolingalgorithm} only require that the initial state lies in a sufficiently low-energy subspace; in some instances this can lead to high overlap with the ground state, but this is certainly not guaranteed in general, assuming no collapse of complexity classes. With an appropriate snaking of the 2D lattice, a faulty MPS approximation of the groundstate can thus be transferred to the quantum processor.

\subsection{Entangled Logical Qubits}
\label{sec:logical}
Going beyond groundstates of local Hamiltonians, in this section we challenge CVD on states whose entanglement is distributed such that it is not removable by single local operations. Consider the example of a Bell pair distributed on far away sites. Its state complexity is proportional to the distance of the entangled qubits, as the state can only be disentangled by first swapping the pair to neighboring sites and subsequently applying a disentangling gate, a CNOT gate for instance. Swapping the Bell pair closer together only decreases the entanglement entropy of the left- and rightmost bond. Thus, there is hope that CVD, which only looks at pairs of two neighboring sites at a time, can find a good approximation of this state.

In order to make the example more interesting, we additionally smear out the entanglement among many qubits by encoding the Bell pair in a stabilizer code. There, local operations should not alter the state, if stabilizer measurements are carried out fast enough. We thus expect that a single layer of unitaries cannot significantly disentangle the state alone, but there will be a minimal necessary depth to be able to concentrate the entanglement entropy into two neighboring qubits. We consider two examples of stabilizer codes with distances three and five, in particular the [5,1,3] code and the [11,1,5] code defined by the symmetric subspace of the stabilizers displayed in Table~\ref{tab:stabilizers}. Our goal is to disentangle the logical Bell pair $\frac{1}{\sqrt{2}} \left( \ket{01}_L - \ket{10}_L \right)$ that defines a 10-qubit, or 22-qubit quantum state, respectively, shown in Fig.~\ref{fig:QEC} (d).

We plot the results for both, the [5,1,3] and the [11,1,5] code for separated and interlaced logical Bell pairs in Fig.~\ref{fig:QEC}. While the separated states can be prepared exactly (up to numerical precision) with approximately $\frac{n}{2}$ layers, disentangling the interlaced logical qubits needs a deeper circuit to properly swap the entanglement to neighboring sites. Although the optimizer eventually reaches overlap errors of below $10^{-3}$ per site, the interlaced [11,1,5] Bell pair is not exactly prepared, but seems to be stuck in a local minimum. The evolution of the tail weights behaves similar in both codes, while the decrease of the tail weights in the [11,1,5] code are evidently subject to fluctuations. It appears that CVD, being a strictly local disentangling strategy, is having trouble finding the correct order of swapping back the interlaced qubits before removing the non-local entanglement as in the separated case. The attempt to directly remove entanglement from the start, i.e. without swapping qubits around first, leads the optimizer into a convergence barrier.

\section{Conclusion}
We present classical variational disentanglement, or CVD, a classical mapping of MPS representations to quantum circuits for state preparation. The classically efficient representation of MPS using local information is utilized to optimize a parametrized circuit to minimize the entanglement entropy for bipartitions defined by each bond of the MPS. We prove guarantees for classical efficiency assuring trainability and bounds on the maximal bond dimension and the error of the output state and showcase the algorithm on a number of examples.

We expect CVD to be a flexible pre-processing routine for the simulation of quench dynamics, where the initial state is taken as the ground state of a gapped, local Hamiltonian. Since the initial state preparation is not the quantumly hard part, it should be processed classically. Also generic ground state algorithms like the variational quantum eigensolver, imaginary time evolution or cooling protocols benefit from an educated guess, which CVD provides. CVD also improves the access to entangled bases for sample efficient measurements \cite{eckstein2024shotnoisereductionlatticehamiltonians}, as well as for tomography of low-rank states \cite{Huang_2024} and processes \cite{Mansuroglu_2024}, which are further known ways of connecting classical and quantum information processing. CVD is not restricted to disentangling only local correlations, but can in practice become trapped in local minima that yield good approximations but neglect longer-range correlations. Such cases may be alleviated by blocking sites before optimization, which offers a natural extension of the method to more strongly correlated systems.

The idea of disentangling tensor networks with local gates reaches beyond state preparation. Interesting generalizations include classical compilation of circuits for time evolution by disentanglement of matrix product operators (MPOs) going beyond Trotterization-based classical compilation schemes like \cite{Mc_Keever_2023, Mansuroglu_2023, Mansuroglu_2023b} or sequential preparation \cite{Nibbi_2024}. Similarly, matrix product functions (MPFs) have been successfully applied for machine learning tasks \cite{latorre2005imagecompressionentanglement, stoudenmire2017supervisedlearningquantuminspiredtensor} and thus their compilation into circuits (potentially including non-unitary structures) provide classical guesses for quantum machine learning models boosting trainability with CVD as a systematic segue past barren plateaus and towards narrow gorges. 

\begin{acknowledgments}
    This research was funded in parts by the Austrian Science Fund (FWF) via the Cluster of Excellence ``quantA'' (\href{https://doi.org/10.55776/COE1}{10.55776/COE1}) and via the Special Research Programme ``BeyondC: Quantum Information Systems Beyond Classical Capabilities'' (\href{https://doi.org/10.55776/F71}{10.55776/F71}), by the European Union -- NextGenerationEU, and by the European Union’s Horizon 2020 research and innovation programme through Grant No.\ 863476 (ERC-CoG SEQUAM).
\end{acknowledgments}

\bibliographystyle{apsrev4-2}
\bibliography{BiblioR}

\begin{thebibliography}{48}%
\makeatletter
\providecommand \@ifxundefined [1]{%
 \@ifx{#1\undefined}
}%
\providecommand \@ifnum [1]{%
 \ifnum #1\expandafter \@firstoftwo
 \else \expandafter \@secondoftwo
 \fi
}%
\providecommand \@ifx [1]{%
 \ifx #1\expandafter \@firstoftwo
 \else \expandafter \@secondoftwo
 \fi
}%
\providecommand \natexlab [1]{#1}%
\providecommand \enquote  [1]{``#1''}%
\providecommand \bibnamefont  [1]{#1}%
\providecommand \bibfnamefont [1]{#1}%
\providecommand \citenamefont [1]{#1}%
\providecommand \href@noop [0]{\@secondoftwo}%
\providecommand \href [0]{\begingroup \@sanitize@url \@href}%
\providecommand \@href[1]{\@@startlink{#1}\@@href}%
\providecommand \@@href[1]{\endgroup#1\@@endlink}%
\providecommand \@sanitize@url [0]{\catcode `\\12\catcode `\$12\catcode
  `\&12\catcode `\#12\catcode `\^12\catcode `\_12\catcode `\%12\relax}%
\providecommand \@@startlink[1]{}%
\providecommand \@@endlink[0]{}%
\providecommand \url  [0]{\begingroup\@sanitize@url \@url }%
\providecommand \@url [1]{\endgroup\@href {#1}{\urlprefix }}%
\providecommand \urlprefix  [0]{URL }%
\providecommand \Eprint [0]{\href }%
\providecommand \doibase [0]{https://doi.org/}%
\providecommand \selectlanguage [0]{\@gobble}%
\providecommand \bibinfo  [0]{\@secondoftwo}%
\providecommand \bibfield  [0]{\@secondoftwo}%
\providecommand \translation [1]{[#1]}%
\providecommand \BibitemOpen [0]{}%
\providecommand \bibitemStop [0]{}%
\providecommand \bibitemNoStop [0]{.\EOS\space}%
\providecommand \EOS [0]{\spacefactor3000\relax}%
\providecommand \BibitemShut  [1]{\csname bibitem#1\endcsname}%
\let\auto@bib@innerbib\@empty
\bibitem [{\citenamefont {Marti}\ \emph {et~al.}(2025)\citenamefont {Marti},
  \citenamefont {Mansuroglu},\ and\ \citenamefont
  {Hartmann}}]{marti2025efficientquantumcoolingalgorithm}%
  \BibitemOpen
  \bibfield  {author} {\bibinfo {author} {\bibfnamefont {L.}~\bibnamefont
  {Marti}}, \bibinfo {author} {\bibfnamefont {R.}~\bibnamefont {Mansuroglu}},\
  and\ \bibinfo {author} {\bibfnamefont {M.~J.}\ \bibnamefont {Hartmann}},\
  }\href {https://doi.org/10.22331/q-2025-02-18-1635} {\bibfield  {journal}
  {\bibinfo  {journal} {Quantum}\ }\textbf {\bibinfo {volume} {9}},\ \bibinfo
  {pages} {1635} (\bibinfo {year} {2025})}\BibitemShut {NoStop}%
\bibitem [{\citenamefont {McClean}\ \emph {et~al.}(2018)\citenamefont
  {McClean}, \citenamefont {Boixo}, \citenamefont {Smelyanskiy}, \citenamefont
  {Babbush},\ and\ \citenamefont {Neven}}]{McClean_2018}%
  \BibitemOpen
  \bibfield  {author} {\bibinfo {author} {\bibfnamefont {J.~R.}\ \bibnamefont
  {McClean}}, \bibinfo {author} {\bibfnamefont {S.}~\bibnamefont {Boixo}},
  \bibinfo {author} {\bibfnamefont {V.~N.}\ \bibnamefont {Smelyanskiy}},
  \bibinfo {author} {\bibfnamefont {R.}~\bibnamefont {Babbush}},\ and\ \bibinfo
  {author} {\bibfnamefont {H.}~\bibnamefont {Neven}},\ }\bibfield  {journal}
  {\bibinfo  {journal} {Nature Communications}\ }\textbf {\bibinfo {volume}
  {9}},\ \href {https://doi.org/10.1038/s41467-018-07090-4}
  {10.1038/s41467-018-07090-4} (\bibinfo {year} {2018})\BibitemShut {NoStop}%
\bibitem [{\citenamefont {Verstraete}\ and\ \citenamefont
  {Cirac}(2006)}]{Verstraete_2006}%
  \BibitemOpen
  \bibfield  {author} {\bibinfo {author} {\bibfnamefont {F.}~\bibnamefont
  {Verstraete}}\ and\ \bibinfo {author} {\bibfnamefont {J.~I.}\ \bibnamefont
  {Cirac}},\ }\bibfield  {journal} {\bibinfo  {journal} {Physical Review B}\
  }\textbf {\bibinfo {volume} {73}},\ \href
  {https://doi.org/10.1103/physrevb.73.094423} {10.1103/physrevb.73.094423}
  (\bibinfo {year} {2006})\BibitemShut {NoStop}%
\bibitem [{\citenamefont {Hastings}(2007)}]{Hastings_2007}%
  \BibitemOpen
  \bibfield  {author} {\bibinfo {author} {\bibfnamefont {M.~B.}\ \bibnamefont
  {Hastings}},\ }\href {https://doi.org/10.1088/1742-5468/2007/08/p08024}
  {\bibfield  {journal} {\bibinfo  {journal} {Journal of Statistical Mechanics:
  Theory and Experiment}\ }\textbf {\bibinfo {volume} {2007}},\ \bibinfo
  {pages} {P08024–P08024} (\bibinfo {year} {2007})}\BibitemShut {NoStop}%
\bibitem [{\citenamefont {Schuch}\ \emph
  {et~al.}(2008{\natexlab{a}})\citenamefont {Schuch}, \citenamefont {Wolf},
  \citenamefont {Vollbrecht},\ and\ \citenamefont {Cirac}}]{Schuch_2008b}%
  \BibitemOpen
  \bibfield  {author} {\bibinfo {author} {\bibfnamefont {N.}~\bibnamefont
  {Schuch}}, \bibinfo {author} {\bibfnamefont {M.~M.}\ \bibnamefont {Wolf}},
  \bibinfo {author} {\bibfnamefont {K.~G.~H.}\ \bibnamefont {Vollbrecht}},\
  and\ \bibinfo {author} {\bibfnamefont {J.~I.}\ \bibnamefont {Cirac}},\ }\href
  {https://doi.org/10.1088/1367-2630/10/3/033032} {\bibfield  {journal}
  {\bibinfo  {journal} {New Journal of Physics}\ }\textbf {\bibinfo {volume}
  {10}},\ \bibinfo {pages} {033032} (\bibinfo {year}
  {2008}{\natexlab{a}})}\BibitemShut {NoStop}%
\bibitem [{\citenamefont {Will}\ \emph {et~al.}(2025)\citenamefont {Will},
  \citenamefont {Cochran}, \citenamefont {Rosenberg}, \citenamefont {Jobst},
  \citenamefont {Eassa}, \citenamefont {Roushan}, \citenamefont {Knap},
  \citenamefont {Gammon-Smith},\ and\ \citenamefont
  {Pollmann}}]{will2025probingnonequilibriumtopologicalorder}%
  \BibitemOpen
  \bibfield  {author} {\bibinfo {author} {\bibfnamefont {M.}~\bibnamefont
  {Will}}, \bibinfo {author} {\bibfnamefont {T.~A.}\ \bibnamefont {Cochran}},
  \bibinfo {author} {\bibfnamefont {E.}~\bibnamefont {Rosenberg}}, \bibinfo
  {author} {\bibfnamefont {B.}~\bibnamefont {Jobst}}, \bibinfo {author}
  {\bibfnamefont {N.~M.}\ \bibnamefont {Eassa}}, \bibinfo {author}
  {\bibfnamefont {P.}~\bibnamefont {Roushan}}, \bibinfo {author} {\bibfnamefont
  {M.}~\bibnamefont {Knap}}, \bibinfo {author} {\bibfnamefont {A.}~\bibnamefont
  {Gammon-Smith}},\ and\ \bibinfo {author} {\bibfnamefont {F.}~\bibnamefont
  {Pollmann}},\ }\href {https://arxiv.org/abs/2501.18461} {\bibinfo {title}
  {Probing non-equilibrium topological order on a quantum processor}} (\bibinfo
  {year} {2025}),\ \Eprint {https://arxiv.org/abs/2501.18461} {arXiv:2501.18461
  [quant-ph]} \BibitemShut {NoStop}%
\bibitem [{\citenamefont {Evered}\ \emph {et~al.}(2025)\citenamefont {Evered},
  \citenamefont {Kalinowski}, \citenamefont {Geim}, \citenamefont {Manovitz},
  \citenamefont {Bluvstein}, \citenamefont {Li}, \citenamefont {Maskara},
  \citenamefont {Zhou}, \citenamefont {Ebadi}, \citenamefont {Xu},
  \citenamefont {Campo}, \citenamefont {Cain}, \citenamefont {Ostermann},
  \citenamefont {Yelin}, \citenamefont {Sachdev}, \citenamefont {Greiner},
  \citenamefont {Vuletić},\ and\ \citenamefont
  {Lukin}}]{evered2025probingtopologicalmatterfermion}%
  \BibitemOpen
  \bibfield  {author} {\bibinfo {author} {\bibfnamefont {S.~J.}\ \bibnamefont
  {Evered}}, \bibinfo {author} {\bibfnamefont {M.}~\bibnamefont {Kalinowski}},
  \bibinfo {author} {\bibfnamefont {A.~A.}\ \bibnamefont {Geim}}, \bibinfo
  {author} {\bibfnamefont {T.}~\bibnamefont {Manovitz}}, \bibinfo {author}
  {\bibfnamefont {D.}~\bibnamefont {Bluvstein}}, \bibinfo {author}
  {\bibfnamefont {S.~H.}\ \bibnamefont {Li}}, \bibinfo {author} {\bibfnamefont
  {N.}~\bibnamefont {Maskara}}, \bibinfo {author} {\bibfnamefont
  {H.}~\bibnamefont {Zhou}}, \bibinfo {author} {\bibfnamefont {S.}~\bibnamefont
  {Ebadi}}, \bibinfo {author} {\bibfnamefont {M.}~\bibnamefont {Xu}}, \bibinfo
  {author} {\bibfnamefont {J.}~\bibnamefont {Campo}}, \bibinfo {author}
  {\bibfnamefont {M.}~\bibnamefont {Cain}}, \bibinfo {author} {\bibfnamefont
  {S.}~\bibnamefont {Ostermann}}, \bibinfo {author} {\bibfnamefont {S.~F.}\
  \bibnamefont {Yelin}}, \bibinfo {author} {\bibfnamefont {S.}~\bibnamefont
  {Sachdev}}, \bibinfo {author} {\bibfnamefont {M.}~\bibnamefont {Greiner}},
  \bibinfo {author} {\bibfnamefont {V.}~\bibnamefont {Vuletić}},\ and\
  \bibinfo {author} {\bibfnamefont {M.~D.}\ \bibnamefont {Lukin}},\ }\href
  {https://arxiv.org/abs/2501.18554} {\bibinfo {title} {Probing topological
  matter and fermion dynamics on a neutral-atom quantum computer}} (\bibinfo
  {year} {2025}),\ \Eprint {https://arxiv.org/abs/2501.18554} {arXiv:2501.18554
  [quant-ph]} \BibitemShut {NoStop}%
\bibitem [{\citenamefont {Schön}\ \emph {et~al.}(2005)\citenamefont {Schön},
  \citenamefont {Solano}, \citenamefont {Verstraete}, \citenamefont {Cirac},\
  and\ \citenamefont {Wolf}}]{Sch_n_2005}%
  \BibitemOpen
  \bibfield  {author} {\bibinfo {author} {\bibfnamefont {C.}~\bibnamefont
  {Schön}}, \bibinfo {author} {\bibfnamefont {E.}~\bibnamefont {Solano}},
  \bibinfo {author} {\bibfnamefont {F.}~\bibnamefont {Verstraete}}, \bibinfo
  {author} {\bibfnamefont {J.~I.}\ \bibnamefont {Cirac}},\ and\ \bibinfo
  {author} {\bibfnamefont {M.~M.}\ \bibnamefont {Wolf}},\ }\bibfield  {journal}
  {\bibinfo  {journal} {Physical Review Letters}\ }\textbf {\bibinfo {volume}
  {95}},\ \href {https://doi.org/10.1103/physrevlett.95.110503}
  {10.1103/physrevlett.95.110503} (\bibinfo {year} {2005})\BibitemShut
  {NoStop}%
\bibitem [{\citenamefont {Malz}\ \emph {et~al.}(2024)\citenamefont {Malz},
  \citenamefont {Styliaris}, \citenamefont {Wei},\ and\ \citenamefont
  {Cirac}}]{Malz_2024}%
  \BibitemOpen
  \bibfield  {author} {\bibinfo {author} {\bibfnamefont {D.}~\bibnamefont
  {Malz}}, \bibinfo {author} {\bibfnamefont {G.}~\bibnamefont {Styliaris}},
  \bibinfo {author} {\bibfnamefont {Z.-Y.}\ \bibnamefont {Wei}},\ and\ \bibinfo
  {author} {\bibfnamefont {J.~I.}\ \bibnamefont {Cirac}},\ }\bibfield
  {journal} {\bibinfo  {journal} {Physical Review Letters}\ }\textbf {\bibinfo
  {volume} {132}},\ \href {https://doi.org/10.1103/physrevlett.132.040404}
  {10.1103/physrevlett.132.040404} (\bibinfo {year} {2024})\BibitemShut
  {NoStop}%
\bibitem [{\citenamefont {Wei}\ and\ \citenamefont
  {Malz}(2025)}]{wei2025statepreparationparallelsequentialcircuits}%
  \BibitemOpen
  \bibfield  {author} {\bibinfo {author} {\bibfnamefont {Z.-Y.}\ \bibnamefont
  {Wei}}\ and\ \bibinfo {author} {\bibfnamefont {D.}~\bibnamefont {Malz}},\
  }\href {https://arxiv.org/abs/2503.14645} {\bibinfo {title} {State
  preparation with parallel-sequential circuits}} (\bibinfo {year} {2025}),\
  \Eprint {https://arxiv.org/abs/2503.14645} {arXiv:2503.14645 [quant-ph]}
  \BibitemShut {NoStop}%
\bibitem [{\citenamefont {Smith}\ \emph {et~al.}(2024)\citenamefont {Smith},
  \citenamefont {Khan}, \citenamefont {Clark}, \citenamefont {Girvin},\ and\
  \citenamefont {Wei}}]{Smith24}%
  \BibitemOpen
  \bibfield  {author} {\bibinfo {author} {\bibfnamefont {K.~C.}\ \bibnamefont
  {Smith}}, \bibinfo {author} {\bibfnamefont {A.}~\bibnamefont {Khan}},
  \bibinfo {author} {\bibfnamefont {B.~K.}\ \bibnamefont {Clark}}, \bibinfo
  {author} {\bibfnamefont {S.}~\bibnamefont {Girvin}},\ and\ \bibinfo {author}
  {\bibfnamefont {T.-C.}\ \bibnamefont {Wei}},\ }\href
  {https://doi.org/10.1103/PRXQuantum.5.030344} {\bibfield  {journal} {\bibinfo
   {journal} {PRX Quantum}\ }\textbf {\bibinfo {volume} {5}},\ \bibinfo {pages}
  {030344} (\bibinfo {year} {2024})}\BibitemShut {NoStop}%
\bibitem [{\citenamefont {Ran}(2020)}]{Ran_2020}%
  \BibitemOpen
  \bibfield  {author} {\bibinfo {author} {\bibfnamefont {S.-J.}\ \bibnamefont
  {Ran}},\ }\bibfield  {journal} {\bibinfo  {journal} {Physical Review A}\
  }\textbf {\bibinfo {volume} {101}},\ \href
  {https://doi.org/10.1103/physreva.101.032310} {10.1103/physreva.101.032310}
  (\bibinfo {year} {2020})\BibitemShut {NoStop}%
\bibitem [{\citenamefont {Rudolph}\ \emph {et~al.}(2023)\citenamefont
  {Rudolph}, \citenamefont {Chen}, \citenamefont {Miller}, \citenamefont
  {Acharya},\ and\ \citenamefont {Perdomo-Ortiz}}]{Rudolph_2024}%
  \BibitemOpen
  \bibfield  {author} {\bibinfo {author} {\bibfnamefont {M.~S.}\ \bibnamefont
  {Rudolph}}, \bibinfo {author} {\bibfnamefont {J.}~\bibnamefont {Chen}},
  \bibinfo {author} {\bibfnamefont {J.}~\bibnamefont {Miller}}, \bibinfo
  {author} {\bibfnamefont {A.}~\bibnamefont {Acharya}},\ and\ \bibinfo {author}
  {\bibfnamefont {A.}~\bibnamefont {Perdomo-Ortiz}},\ }\href
  {https://doi.org/10.1088/2058-9565/ad04e6} {\bibfield  {journal} {\bibinfo
  {journal} {Quantum Science and Technology}\ }\textbf {\bibinfo {volume}
  {9}},\ \bibinfo {pages} {015012} (\bibinfo {year} {2023})}\BibitemShut
  {NoStop}%
\bibitem [{\citenamefont {Jobst}\ \emph {et~al.}(2024)\citenamefont {Jobst},
  \citenamefont {Shen}, \citenamefont {Riofrío}, \citenamefont {Shishenina},\
  and\ \citenamefont {Pollmann}}]{Jobst_2024}%
  \BibitemOpen
  \bibfield  {author} {\bibinfo {author} {\bibfnamefont {B.}~\bibnamefont
  {Jobst}}, \bibinfo {author} {\bibfnamefont {K.}~\bibnamefont {Shen}},
  \bibinfo {author} {\bibfnamefont {C.~A.}\ \bibnamefont {Riofrío}}, \bibinfo
  {author} {\bibfnamefont {E.}~\bibnamefont {Shishenina}},\ and\ \bibinfo
  {author} {\bibfnamefont {F.}~\bibnamefont {Pollmann}},\ }\href
  {https://doi.org/10.22331/q-2024-12-03-1544} {\bibfield  {journal} {\bibinfo
  {journal} {Quantum}\ }\textbf {\bibinfo {volume} {8}},\ \bibinfo {pages}
  {1544} (\bibinfo {year} {2024})}\BibitemShut {NoStop}%
\bibitem [{\citenamefont
  {Murota}(2025)}]{murota2025adiabaticencodingpretrainedmps}%
  \BibitemOpen
  \bibfield  {author} {\bibinfo {author} {\bibfnamefont {K.}~\bibnamefont
  {Murota}},\ }\href {https://arxiv.org/abs/2504.09250} {\bibinfo {title}
  {Adiabatic encoding of pre-trained mps classifiers into quantum circuits}}
  (\bibinfo {year} {2025}),\ \Eprint {https://arxiv.org/abs/2504.09250}
  {arXiv:2504.09250 [quant-ph]} \BibitemShut {NoStop}%
\bibitem [{\citenamefont {Jaderberg}\ \emph {et~al.}(2025)\citenamefont
  {Jaderberg}, \citenamefont {Pennington}, \citenamefont {Marshall},
  \citenamefont {Anderson}, \citenamefont {Agarwal}, \citenamefont {Lindoy},
  \citenamefont {Rungger}, \citenamefont {Mensa},\ and\ \citenamefont
  {Crain}}]{jaderberg2025variationalpreparationnormalmatrix}%
  \BibitemOpen
  \bibfield  {author} {\bibinfo {author} {\bibfnamefont {B.}~\bibnamefont
  {Jaderberg}}, \bibinfo {author} {\bibfnamefont {G.}~\bibnamefont
  {Pennington}}, \bibinfo {author} {\bibfnamefont {K.~V.}\ \bibnamefont
  {Marshall}}, \bibinfo {author} {\bibfnamefont {L.~W.}\ \bibnamefont
  {Anderson}}, \bibinfo {author} {\bibfnamefont {A.}~\bibnamefont {Agarwal}},
  \bibinfo {author} {\bibfnamefont {L.~P.}\ \bibnamefont {Lindoy}}, \bibinfo
  {author} {\bibfnamefont {I.}~\bibnamefont {Rungger}}, \bibinfo {author}
  {\bibfnamefont {S.}~\bibnamefont {Mensa}},\ and\ \bibinfo {author}
  {\bibfnamefont {J.}~\bibnamefont {Crain}},\ }\href
  {https://arxiv.org/abs/2503.09683} {\bibinfo {title} {Variational preparation
  of normal matrix product states on quantum computers}} (\bibinfo {year}
  {2025}),\ \Eprint {https://arxiv.org/abs/2503.09683} {arXiv:2503.09683
  [quant-ph]} \BibitemShut {NoStop}%
\bibitem [{\citenamefont {Robertson}\ \emph {et~al.}(2024)\citenamefont
  {Robertson}, \citenamefont {Akhriev}, \citenamefont {Vala},\ and\
  \citenamefont {Zhuk}}]{robertson2024approximatequantumcompilingquantum}%
  \BibitemOpen
  \bibfield  {author} {\bibinfo {author} {\bibfnamefont {N.~F.}\ \bibnamefont
  {Robertson}}, \bibinfo {author} {\bibfnamefont {A.}~\bibnamefont {Akhriev}},
  \bibinfo {author} {\bibfnamefont {J.}~\bibnamefont {Vala}},\ and\ \bibinfo
  {author} {\bibfnamefont {S.}~\bibnamefont {Zhuk}},\ }\href
  {https://arxiv.org/abs/2301.08609} {\bibinfo {title} {Approximate quantum
  compiling for quantum simulation: A tensor network based approach}} (\bibinfo
  {year} {2024}),\ \Eprint {https://arxiv.org/abs/2301.08609} {arXiv:2301.08609
  [quant-ph]} \BibitemShut {NoStop}%
\bibitem [{\citenamefont {Lubasch}\ \emph {et~al.}(2020)\citenamefont
  {Lubasch}, \citenamefont {Joo}, \citenamefont {Moinier}, \citenamefont
  {Kiffner},\ and\ \citenamefont {Jaksch}}]{Lubasch_2020}%
  \BibitemOpen
  \bibfield  {author} {\bibinfo {author} {\bibfnamefont {M.}~\bibnamefont
  {Lubasch}}, \bibinfo {author} {\bibfnamefont {J.}~\bibnamefont {Joo}},
  \bibinfo {author} {\bibfnamefont {P.}~\bibnamefont {Moinier}}, \bibinfo
  {author} {\bibfnamefont {M.}~\bibnamefont {Kiffner}},\ and\ \bibinfo {author}
  {\bibfnamefont {D.}~\bibnamefont {Jaksch}},\ }\bibfield  {journal} {\bibinfo
  {journal} {Physical Review A}\ }\textbf {\bibinfo {volume} {101}},\ \href
  {https://doi.org/10.1103/physreva.101.010301} {10.1103/physreva.101.010301}
  (\bibinfo {year} {2020})\BibitemShut {NoStop}%
\bibitem [{\citenamefont {Mc~Keever}\ and\ \citenamefont
  {Lubasch}(2024)}]{Mc_Keever_2024}%
  \BibitemOpen
  \bibfield  {author} {\bibinfo {author} {\bibfnamefont {C.}~\bibnamefont
  {Mc~Keever}}\ and\ \bibinfo {author} {\bibfnamefont {M.}~\bibnamefont
  {Lubasch}},\ }\bibfield  {journal} {\bibinfo  {journal} {PRX Quantum}\
  }\textbf {\bibinfo {volume} {5}},\ \href
  {https://doi.org/10.1103/prxquantum.5.020362} {10.1103/prxquantum.5.020362}
  (\bibinfo {year} {2024})\BibitemShut {NoStop}%
\bibitem [{\citenamefont {Ben-Dov}\ \emph {et~al.}(2024)\citenamefont
  {Ben-Dov}, \citenamefont {Shnaiderov}, \citenamefont {Makmal},\ and\
  \citenamefont {Dalla~Torre}}]{Ben_Dov_2024}%
  \BibitemOpen
  \bibfield  {author} {\bibinfo {author} {\bibfnamefont {M.}~\bibnamefont
  {Ben-Dov}}, \bibinfo {author} {\bibfnamefont {D.}~\bibnamefont {Shnaiderov}},
  \bibinfo {author} {\bibfnamefont {A.}~\bibnamefont {Makmal}},\ and\ \bibinfo
  {author} {\bibfnamefont {E.~G.}\ \bibnamefont {Dalla~Torre}},\ }\bibfield
  {journal} {\bibinfo  {journal} {npj Quantum Information}\ }\textbf {\bibinfo
  {volume} {10}},\ \href {https://doi.org/10.1038/s41534-024-00858-1}
  {10.1038/s41534-024-00858-1} (\bibinfo {year} {2024})\BibitemShut {NoStop}%
\bibitem [{\citenamefont {Khan}\ \emph {et~al.}(2023)\citenamefont {Khan},
  \citenamefont {Clark},\ and\ \citenamefont
  {Tubman}}]{khan2023preoptimizingvariationalquantumeigensolvers}%
  \BibitemOpen
  \bibfield  {author} {\bibinfo {author} {\bibfnamefont {A.}~\bibnamefont
  {Khan}}, \bibinfo {author} {\bibfnamefont {B.~K.}\ \bibnamefont {Clark}},\
  and\ \bibinfo {author} {\bibfnamefont {N.~M.}\ \bibnamefont {Tubman}},\
  }\href {https://arxiv.org/abs/2310.12965} {\bibinfo {title} {Pre-optimizing
  variational quantum eigensolvers with tensor networks}} (\bibinfo {year}
  {2023}),\ \Eprint {https://arxiv.org/abs/2310.12965} {arXiv:2310.12965
  [quant-ph]} \BibitemShut {NoStop}%
\bibitem [{\citenamefont {Melnikov}\ \emph {et~al.}(2023)\citenamefont
  {Melnikov}, \citenamefont {Termanova}, \citenamefont {Dolgov}, \citenamefont
  {Neukart},\ and\ \citenamefont {Perelshtein}}]{Melnikov_2023}%
  \BibitemOpen
  \bibfield  {author} {\bibinfo {author} {\bibfnamefont {A.~A.}\ \bibnamefont
  {Melnikov}}, \bibinfo {author} {\bibfnamefont {A.~A.}\ \bibnamefont
  {Termanova}}, \bibinfo {author} {\bibfnamefont {S.~V.}\ \bibnamefont
  {Dolgov}}, \bibinfo {author} {\bibfnamefont {F.}~\bibnamefont {Neukart}},\
  and\ \bibinfo {author} {\bibfnamefont {M.~R.}\ \bibnamefont {Perelshtein}},\
  }\href {https://doi.org/10.1088/2058-9565/acd9e7} {\bibfield  {journal}
  {\bibinfo  {journal} {Quantum Science and Technology}\ }\textbf {\bibinfo
  {volume} {8}},\ \bibinfo {pages} {035027} (\bibinfo {year}
  {2023})}\BibitemShut {NoStop}%
\bibitem [{\citenamefont {Vidal}(2003)}]{Vidal_2003}%
  \BibitemOpen
  \bibfield  {author} {\bibinfo {author} {\bibfnamefont {G.}~\bibnamefont
  {Vidal}},\ }\bibfield  {journal} {\bibinfo  {journal} {Physical Review
  Letters}\ }\textbf {\bibinfo {volume} {91}},\ \href
  {https://doi.org/10.1103/physrevlett.91.147902}
  {10.1103/physrevlett.91.147902} (\bibinfo {year} {2003})\BibitemShut
  {NoStop}%
\bibitem [{\citenamefont {Schollwöck}(2011)}]{Schollw_ck_2011}%
  \BibitemOpen
  \bibfield  {author} {\bibinfo {author} {\bibfnamefont {U.}~\bibnamefont
  {Schollwöck}},\ }\href {https://doi.org/10.1016/j.aop.2010.09.012}
  {\bibfield  {journal} {\bibinfo  {journal} {Annals of Physics}\ }\textbf
  {\bibinfo {volume} {326}},\ \bibinfo {pages} {96–192} (\bibinfo {year}
  {2011})}\BibitemShut {NoStop}%
\bibitem [{\citenamefont {Kraus}\ and\ \citenamefont
  {Cirac}(2001)}]{Kraus_2001}%
  \BibitemOpen
  \bibfield  {author} {\bibinfo {author} {\bibfnamefont {B.}~\bibnamefont
  {Kraus}}\ and\ \bibinfo {author} {\bibfnamefont {J.~I.}\ \bibnamefont
  {Cirac}},\ }\bibfield  {journal} {\bibinfo  {journal} {Physical Review A}\
  }\textbf {\bibinfo {volume} {63}},\ \href
  {https://doi.org/10.1103/physreva.63.062309} {10.1103/physreva.63.062309}
  (\bibinfo {year} {2001})\BibitemShut {NoStop}%
\bibitem [{\citenamefont {Cerezo}\ \emph {et~al.}(2024)\citenamefont {Cerezo},
  \citenamefont {Larocca}, \citenamefont {García-Martín}, \citenamefont
  {Diaz}, \citenamefont {Braccia}, \citenamefont {Fontana}, \citenamefont
  {Rudolph}, \citenamefont {Bermejo}, \citenamefont {Ijaz}, \citenamefont
  {Thanasilp}, \citenamefont {Anschuetz},\ and\ \citenamefont
  {Holmes}}]{cerezo2024doesprovableabsencebarren}%
  \BibitemOpen
  \bibfield  {author} {\bibinfo {author} {\bibfnamefont {M.}~\bibnamefont
  {Cerezo}}, \bibinfo {author} {\bibfnamefont {M.}~\bibnamefont {Larocca}},
  \bibinfo {author} {\bibfnamefont {D.}~\bibnamefont {García-Martín}},
  \bibinfo {author} {\bibfnamefont {N.~L.}\ \bibnamefont {Diaz}}, \bibinfo
  {author} {\bibfnamefont {P.}~\bibnamefont {Braccia}}, \bibinfo {author}
  {\bibfnamefont {E.}~\bibnamefont {Fontana}}, \bibinfo {author} {\bibfnamefont
  {M.~S.}\ \bibnamefont {Rudolph}}, \bibinfo {author} {\bibfnamefont
  {P.}~\bibnamefont {Bermejo}}, \bibinfo {author} {\bibfnamefont
  {A.}~\bibnamefont {Ijaz}}, \bibinfo {author} {\bibfnamefont {S.}~\bibnamefont
  {Thanasilp}}, \bibinfo {author} {\bibfnamefont {E.~R.}\ \bibnamefont
  {Anschuetz}},\ and\ \bibinfo {author} {\bibfnamefont {Z.}~\bibnamefont
  {Holmes}},\ }\href {https://arxiv.org/abs/2312.09121} {\bibinfo {title} {Does
  provable absence of barren plateaus imply classical simulability? or, why we
  need to rethink variational quantum computing}} (\bibinfo {year} {2024}),\
  \Eprint {https://arxiv.org/abs/2312.09121} {arXiv:2312.09121 [quant-ph]}
  \BibitemShut {NoStop}%
\bibitem [{\citenamefont {Cigna}\ and\ \citenamefont
  {Schilling}(2025)}]{Ludvik}%
  \BibitemOpen
  \bibfield  {author} {\bibinfo {author} {\bibfnamefont {L.}~\bibnamefont
  {Cigna}}\ and\ \bibinfo {author} {\bibfnamefont {C.}~\bibnamefont
  {Schilling}},\ }\href@noop {} {} (\bibinfo {year} {2025}),\ \bibinfo {note}
  {{Private Communication}}\BibitemShut {NoStop}%
\bibitem [{\citenamefont {Fishman}\ \emph {et~al.}(2022)\citenamefont
  {Fishman}, \citenamefont {White},\ and\ \citenamefont
  {Stoudenmire}}]{ITensor}%
  \BibitemOpen
  \bibfield  {author} {\bibinfo {author} {\bibfnamefont {M.}~\bibnamefont
  {Fishman}}, \bibinfo {author} {\bibfnamefont {S.~R.}\ \bibnamefont {White}},\
  and\ \bibinfo {author} {\bibfnamefont {E.~M.}\ \bibnamefont {Stoudenmire}},\
  }\href {https://doi.org/10.21468/SciPostPhysCodeb.4} {\bibfield  {journal}
  {\bibinfo  {journal} {SciPost Phys. Codebases}\ ,\ \bibinfo {pages} {4}}
  (\bibinfo {year} {2022})}\BibitemShut {NoStop}%
\bibitem [{\citenamefont {Mansuroglu}\ and\ \citenamefont
  {Schuch}(2025)}]{CVD_repo}%
  \BibitemOpen
  \bibfield  {author} {\bibinfo {author} {\bibfnamefont {R.}~\bibnamefont
  {Mansuroglu}}\ and\ \bibinfo {author} {\bibfnamefont {N.}~\bibnamefont
  {Schuch}},\ }\href {https://doi.org/10.5281/zenodo.15058443} {\bibinfo
  {title} {Classical variational disentanglement}},\ \bibinfo {howpublished}
  {Software on Zenodo} (\bibinfo {year} {2025}),\ \bibinfo {note} {available
  online at: \url{https://doi.org/10.5281/zenodo.15058443}, last accessed on
  24.04.2025}\BibitemShut {NoStop}%
\bibitem [{\citenamefont {Laflamme}\ \emph {et~al.}(1996)\citenamefont
  {Laflamme}, \citenamefont {Miquel}, \citenamefont {Paz},\ and\ \citenamefont
  {Zurek}}]{laflamme1996perfectquantumerrorcorrection}%
  \BibitemOpen
  \bibfield  {author} {\bibinfo {author} {\bibfnamefont {R.}~\bibnamefont
  {Laflamme}}, \bibinfo {author} {\bibfnamefont {C.}~\bibnamefont {Miquel}},
  \bibinfo {author} {\bibfnamefont {J.~P.}\ \bibnamefont {Paz}},\ and\ \bibinfo
  {author} {\bibfnamefont {W.~H.}\ \bibnamefont {Zurek}},\ }\href
  {https://arxiv.org/abs/quant-ph/9602019} {\bibinfo {title} {Perfect quantum
  error correction code}} (\bibinfo {year} {1996}),\ \Eprint
  {https://arxiv.org/abs/quant-ph/9602019} {arXiv:quant-ph/9602019 [quant-ph]}
  \BibitemShut {NoStop}%
\bibitem [{\citenamefont {Calderbank}\ \emph {et~al.}(1997)\citenamefont
  {Calderbank}, \citenamefont {Rains}, \citenamefont {Shor},\ and\
  \citenamefont {Sloane}}]{calderbank1997quantumerrorcorrectioncodes}%
  \BibitemOpen
  \bibfield  {author} {\bibinfo {author} {\bibfnamefont {A.~R.}\ \bibnamefont
  {Calderbank}}, \bibinfo {author} {\bibfnamefont {E.~M.}\ \bibnamefont
  {Rains}}, \bibinfo {author} {\bibfnamefont {P.~W.}\ \bibnamefont {Shor}},\
  and\ \bibinfo {author} {\bibfnamefont {N.~J.~A.}\ \bibnamefont {Sloane}},\
  }\href {https://arxiv.org/abs/quant-ph/9608006} {\bibinfo {title} {Quantum
  error correction via codes over gf(4)}} (\bibinfo {year} {1997}),\ \Eprint
  {https://arxiv.org/abs/quant-ph/9608006} {arXiv:quant-ph/9608006 [quant-ph]}
  \BibitemShut {NoStop}%
\bibitem [{\citenamefont {Schuch}\ and\ \citenamefont
  {Verstraete}(2009)}]{Schuch_2009}%
  \BibitemOpen
  \bibfield  {author} {\bibinfo {author} {\bibfnamefont {N.}~\bibnamefont
  {Schuch}}\ and\ \bibinfo {author} {\bibfnamefont {F.}~\bibnamefont
  {Verstraete}},\ }\href {https://doi.org/10.1038/nphys1370} {\bibfield
  {journal} {\bibinfo  {journal} {Nature Physics}\ }\textbf {\bibinfo {volume}
  {5}},\ \bibinfo {pages} {732–735} (\bibinfo {year} {2009})}\BibitemShut
  {NoStop}%
\bibitem [{\citenamefont {White}\ and\ \citenamefont
  {Stoudenmire}(2019)}]{White_2019}%
  \BibitemOpen
  \bibfield  {author} {\bibinfo {author} {\bibfnamefont {S.~R.}\ \bibnamefont
  {White}}\ and\ \bibinfo {author} {\bibfnamefont {E.~M.}\ \bibnamefont
  {Stoudenmire}},\ }\bibfield  {journal} {\bibinfo  {journal} {Physical Review
  B}\ }\textbf {\bibinfo {volume} {99}},\ \href
  {https://doi.org/10.1103/physrevb.99.081110} {10.1103/physrevb.99.081110}
  (\bibinfo {year} {2019})\BibitemShut {NoStop}%
\bibitem [{\citenamefont {González-García}\ \emph {et~al.}(2022)\citenamefont
  {González-García}, \citenamefont {Trivedi},\ and\ \citenamefont
  {Cirac}}]{Gonz_lez_Garc_a_2022}%
  \BibitemOpen
  \bibfield  {author} {\bibinfo {author} {\bibfnamefont {G.}~\bibnamefont
  {González-García}}, \bibinfo {author} {\bibfnamefont {R.}~\bibnamefont
  {Trivedi}},\ and\ \bibinfo {author} {\bibfnamefont {J.~I.}\ \bibnamefont
  {Cirac}},\ }\bibfield  {journal} {\bibinfo  {journal} {PRX Quantum}\ }\textbf
  {\bibinfo {volume} {3}},\ \href {https://doi.org/10.1103/prxquantum.3.040326}
  {10.1103/prxquantum.3.040326} (\bibinfo {year} {2022})\BibitemShut {NoStop}%
\bibitem [{\citenamefont {Gharibian}\ and\ \citenamefont
  {Le~Gall}(2023)}]{Gharibian_2023}%
  \BibitemOpen
  \bibfield  {author} {\bibinfo {author} {\bibfnamefont {S.}~\bibnamefont
  {Gharibian}}\ and\ \bibinfo {author} {\bibfnamefont {F.}~\bibnamefont
  {Le~Gall}},\ }\href {https://doi.org/10.1137/22m1513721} {\bibfield
  {journal} {\bibinfo  {journal} {SIAM Journal on Computing}\ }\textbf
  {\bibinfo {volume} {52}},\ \bibinfo {pages} {1009–1038} (\bibinfo {year}
  {2023})}\BibitemShut {NoStop}%
\bibitem [{\citenamefont {Cade}\ \emph {et~al.}(2020)\citenamefont {Cade},
  \citenamefont {Mineh}, \citenamefont {Montanaro},\ and\ \citenamefont
  {Stanisic}}]{Cade2020}%
  \BibitemOpen
  \bibfield  {author} {\bibinfo {author} {\bibfnamefont {C.}~\bibnamefont
  {Cade}}, \bibinfo {author} {\bibfnamefont {L.}~\bibnamefont {Mineh}},
  \bibinfo {author} {\bibfnamefont {A.}~\bibnamefont {Montanaro}},\ and\
  \bibinfo {author} {\bibfnamefont {S.}~\bibnamefont {Stanisic}},\ }\href
  {https://doi.org/10.1103/PhysRevB.102.235122} {\bibfield  {journal} {\bibinfo
   {journal} {Phys. Rev. B}\ }\textbf {\bibinfo {volume} {102}},\ \bibinfo
  {pages} {235122} (\bibinfo {year} {2020})}\BibitemShut {NoStop}%
\bibitem [{\citenamefont {Eckstein}\ \emph {et~al.}(2024)\citenamefont
  {Eckstein}, \citenamefont {Mansuroglu}, \citenamefont {Wolf}, \citenamefont
  {Nützel}, \citenamefont {Tasler}, \citenamefont {Kliesch},\ and\
  \citenamefont
  {Hartmann}}]{eckstein2024shotnoisereductionlatticehamiltonians}%
  \BibitemOpen
  \bibfield  {author} {\bibinfo {author} {\bibfnamefont {T.}~\bibnamefont
  {Eckstein}}, \bibinfo {author} {\bibfnamefont {R.}~\bibnamefont
  {Mansuroglu}}, \bibinfo {author} {\bibfnamefont {S.}~\bibnamefont {Wolf}},
  \bibinfo {author} {\bibfnamefont {L.}~\bibnamefont {Nützel}}, \bibinfo
  {author} {\bibfnamefont {S.}~\bibnamefont {Tasler}}, \bibinfo {author}
  {\bibfnamefont {M.}~\bibnamefont {Kliesch}},\ and\ \bibinfo {author}
  {\bibfnamefont {M.~J.}\ \bibnamefont {Hartmann}},\ }\href
  {https://arxiv.org/abs/2410.21251} {\bibinfo {title} {Shot-noise reduction
  for lattice hamiltonians}} (\bibinfo {year} {2024}),\ \Eprint
  {https://arxiv.org/abs/2410.21251} {arXiv:2410.21251 [quant-ph]} \BibitemShut
  {NoStop}%
\bibitem [{\citenamefont {Huang}\ \emph {et~al.}(2024)\citenamefont {Huang},
  \citenamefont {Liu}, \citenamefont {Broughton}, \citenamefont {Kim},
  \citenamefont {Anshu}, \citenamefont {Landau},\ and\ \citenamefont
  {McClean}}]{Huang_2024}%
  \BibitemOpen
  \bibfield  {author} {\bibinfo {author} {\bibfnamefont {H.-Y.}\ \bibnamefont
  {Huang}}, \bibinfo {author} {\bibfnamefont {Y.}~\bibnamefont {Liu}}, \bibinfo
  {author} {\bibfnamefont {M.}~\bibnamefont {Broughton}}, \bibinfo {author}
  {\bibfnamefont {I.}~\bibnamefont {Kim}}, \bibinfo {author} {\bibfnamefont
  {A.}~\bibnamefont {Anshu}}, \bibinfo {author} {\bibfnamefont
  {Z.}~\bibnamefont {Landau}},\ and\ \bibinfo {author} {\bibfnamefont {J.~R.}\
  \bibnamefont {McClean}},\ }in\ \href
  {https://doi.org/10.1145/3618260.3649722} {\emph {\bibinfo {booktitle}
  {Proceedings of the 56th Annual ACM Symposium on Theory of Computing}}},\
  \bibinfo {series and number} {STOC ’24}\ (\bibinfo  {publisher} {ACM},\
  \bibinfo {year} {2024})\ p.\ \bibinfo {pages} {1343–1351}\BibitemShut
  {NoStop}%
\bibitem [{\citenamefont {Mansuroglu}\ \emph {et~al.}(2024)\citenamefont
  {Mansuroglu}, \citenamefont {Adil}, \citenamefont {Hartmann}, \citenamefont
  {Holmes},\ and\ \citenamefont {Sornborger}}]{Mansuroglu_2024}%
  \BibitemOpen
  \bibfield  {author} {\bibinfo {author} {\bibfnamefont {R.}~\bibnamefont
  {Mansuroglu}}, \bibinfo {author} {\bibfnamefont {A.}~\bibnamefont {Adil}},
  \bibinfo {author} {\bibfnamefont {M.~J.}\ \bibnamefont {Hartmann}}, \bibinfo
  {author} {\bibfnamefont {Z.}~\bibnamefont {Holmes}},\ and\ \bibinfo {author}
  {\bibfnamefont {A.~T.}\ \bibnamefont {Sornborger}},\ }\bibfield  {journal}
  {\bibinfo  {journal} {PRX Quantum}\ }\textbf {\bibinfo {volume} {5}},\ \href
  {https://doi.org/10.1103/prxquantum.5.030306} {10.1103/prxquantum.5.030306}
  (\bibinfo {year} {2024})\BibitemShut {NoStop}%
\bibitem [{\citenamefont {Keever}\ and\ \citenamefont
  {Lubasch}(2023)}]{Mc_Keever_2023}%
  \BibitemOpen
  \bibfield  {author} {\bibinfo {author} {\bibfnamefont {C.~M.}\ \bibnamefont
  {Keever}}\ and\ \bibinfo {author} {\bibfnamefont {M.}~\bibnamefont
  {Lubasch}},\ }\bibfield  {journal} {\bibinfo  {journal} {Physical Review
  Research}\ }\textbf {\bibinfo {volume} {5}},\ \href
  {https://doi.org/10.1103/physrevresearch.5.023146}
  {10.1103/physrevresearch.5.023146} (\bibinfo {year} {2023})\BibitemShut
  {NoStop}%
\bibitem [{\citenamefont {Mansuroglu}\ \emph
  {et~al.}(2023{\natexlab{a}})\citenamefont {Mansuroglu}, \citenamefont
  {Eckstein}, \citenamefont {Nützel}, \citenamefont {Wilkinson},\ and\
  \citenamefont {Hartmann}}]{Mansuroglu_2023}%
  \BibitemOpen
  \bibfield  {author} {\bibinfo {author} {\bibfnamefont {R.}~\bibnamefont
  {Mansuroglu}}, \bibinfo {author} {\bibfnamefont {T.}~\bibnamefont
  {Eckstein}}, \bibinfo {author} {\bibfnamefont {L.}~\bibnamefont {Nützel}},
  \bibinfo {author} {\bibfnamefont {S.~A.}\ \bibnamefont {Wilkinson}},\ and\
  \bibinfo {author} {\bibfnamefont {M.~J.}\ \bibnamefont {Hartmann}},\ }\href
  {https://doi.org/10.1088/2058-9565/acb1d0} {\bibfield  {journal} {\bibinfo
  {journal} {Quantum Science and Technology}\ }\textbf {\bibinfo {volume}
  {8}},\ \bibinfo {pages} {025006} (\bibinfo {year}
  {2023}{\natexlab{a}})}\BibitemShut {NoStop}%
\bibitem [{\citenamefont {Mansuroglu}\ \emph
  {et~al.}(2023{\natexlab{b}})\citenamefont {Mansuroglu}, \citenamefont
  {Fischer},\ and\ \citenamefont {Hartmann}}]{Mansuroglu_2023b}%
  \BibitemOpen
  \bibfield  {author} {\bibinfo {author} {\bibfnamefont {R.}~\bibnamefont
  {Mansuroglu}}, \bibinfo {author} {\bibfnamefont {F.}~\bibnamefont
  {Fischer}},\ and\ \bibinfo {author} {\bibfnamefont {M.~J.}\ \bibnamefont
  {Hartmann}},\ }\bibfield  {journal} {\bibinfo  {journal} {Physical Review
  Research}\ }\textbf {\bibinfo {volume} {5}},\ \href
  {https://doi.org/10.1103/physrevresearch.5.043035}
  {10.1103/physrevresearch.5.043035} (\bibinfo {year}
  {2023}{\natexlab{b}})\BibitemShut {NoStop}%
\bibitem [{\citenamefont {Nibbi}\ and\ \citenamefont
  {Mendl}(2024)}]{Nibbi_2024}%
  \BibitemOpen
  \bibfield  {author} {\bibinfo {author} {\bibfnamefont {M.}~\bibnamefont
  {Nibbi}}\ and\ \bibinfo {author} {\bibfnamefont {C.~B.}\ \bibnamefont
  {Mendl}},\ }\bibfield  {journal} {\bibinfo  {journal} {Physical Review A}\
  }\textbf {\bibinfo {volume} {110}},\ \href
  {https://doi.org/10.1103/physreva.110.042427} {10.1103/physreva.110.042427}
  (\bibinfo {year} {2024})\BibitemShut {NoStop}%
\bibitem [{\citenamefont
  {Latorre}(2005)}]{latorre2005imagecompressionentanglement}%
  \BibitemOpen
  \bibfield  {author} {\bibinfo {author} {\bibfnamefont {J.~I.}\ \bibnamefont
  {Latorre}},\ }\href {https://arxiv.org/abs/quant-ph/0510031} {\bibinfo
  {title} {Image compression and entanglement}} (\bibinfo {year} {2005}),\
  \Eprint {https://arxiv.org/abs/quant-ph/0510031} {arXiv:quant-ph/0510031
  [quant-ph]} \BibitemShut {NoStop}%
\bibitem [{\citenamefont {Stoudenmire}\ and\ \citenamefont
  {Schwab}(2017)}]{stoudenmire2017supervisedlearningquantuminspiredtensor}%
  \BibitemOpen
  \bibfield  {author} {\bibinfo {author} {\bibfnamefont {E.~M.}\ \bibnamefont
  {Stoudenmire}}\ and\ \bibinfo {author} {\bibfnamefont {D.~J.}\ \bibnamefont
  {Schwab}},\ }\href {https://arxiv.org/abs/1605.05775} {\bibinfo {title}
  {Supervised learning with quantum-inspired tensor networks}} (\bibinfo {year}
  {2017}),\ \Eprint {https://arxiv.org/abs/1605.05775} {arXiv:1605.05775
  [stat.ML]} \BibitemShut {NoStop}%
\bibitem [{\citenamefont {Schuch}\ \emph
  {et~al.}(2008{\natexlab{b}})\citenamefont {Schuch}, \citenamefont {Wolf},
  \citenamefont {Verstraete},\ and\ \citenamefont {Cirac}}]{Schuch_2008}%
  \BibitemOpen
  \bibfield  {author} {\bibinfo {author} {\bibfnamefont {N.}~\bibnamefont
  {Schuch}}, \bibinfo {author} {\bibfnamefont {M.~M.}\ \bibnamefont {Wolf}},
  \bibinfo {author} {\bibfnamefont {F.}~\bibnamefont {Verstraete}},\ and\
  \bibinfo {author} {\bibfnamefont {J.~I.}\ \bibnamefont {Cirac}},\ }\bibfield
  {journal} {\bibinfo  {journal} {Physical Review Letters}\ }\textbf {\bibinfo
  {volume} {100}},\ \href {https://doi.org/10.1103/physrevlett.100.030504}
  {10.1103/physrevlett.100.030504} (\bibinfo {year}
  {2008}{\natexlab{b}})\BibitemShut {NoStop}%
\bibitem [{\citenamefont {Chen}\ \emph {et~al.}(2023)\citenamefont {Chen},
  \citenamefont {Shen}, \citenamefont {Lee},\ and\ \citenamefont
  {Yang}}]{Chen_2023}%
  \BibitemOpen
  \bibfield  {author} {\bibinfo {author} {\bibfnamefont {T.}~\bibnamefont
  {Chen}}, \bibinfo {author} {\bibfnamefont {R.}~\bibnamefont {Shen}}, \bibinfo
  {author} {\bibfnamefont {C.~H.}\ \bibnamefont {Lee}},\ and\ \bibinfo {author}
  {\bibfnamefont {B.}~\bibnamefont {Yang}},\ }\bibfield  {journal} {\bibinfo
  {journal} {SciPost Physics}\ }\textbf {\bibinfo {volume} {15}},\ \href
  {https://doi.org/10.21468/scipostphys.15.4.170}
  {10.21468/scipostphys.15.4.170} (\bibinfo {year} {2023})\BibitemShut
  {NoStop}%
\bibitem [{\citenamefont {Chen}\ and\ \citenamefont
  {Byrnes}(2024)}]{Chen_2024}%
  \BibitemOpen
  \bibfield  {author} {\bibinfo {author} {\bibfnamefont {T.}~\bibnamefont
  {Chen}}\ and\ \bibinfo {author} {\bibfnamefont {T.}~\bibnamefont {Byrnes}},\
  }\href {https://doi.org/10.22331/q-2024-12-10-1557} {\bibfield  {journal}
  {\bibinfo  {journal} {Quantum}\ }\textbf {\bibinfo {volume} {8}},\ \bibinfo
  {pages} {1557} (\bibinfo {year} {2024})}\BibitemShut {NoStop}%
\end{thebibliography}%

\onecolumngrid
\appendix
\section{Classical Efficiency of CVD}
In this section, we provide technical details for the proofs of classical efficiency of CVD including a bound on the bond dimension for a given entanglement entropy value, an error bound for the approximation of the prepared state by truncation of MPS and finally a proof of absence of barren plateaus in optimization. Together these results yield good evidence for efficient disentanglement. 

\subsection{Bounds on Bond Dimension -- Proof of Lemma \ref{lem:bond_dim}}
\label{app:bond_dim}
Just as the entanglement entropy is bounded by the bond dimension $D$, vice versa for a given entropy value $S_{\mathcal{A}, \alpha}$, the bond dimension is bounded. We derive a relation for the minimal entropy defined by the squared Schmidt coefficients $\lambda_i^2 =: p_i$, which also satisfy the normalization constraint $\sum_{i=1}^D p_i = 1-p$. The entropy constraint can be reformulated to $\sum_{i=1} p_i^\alpha = e^{S_{\mathcal{A}, \alpha} (1-\alpha)}$. We observe that the gradient of the Lagrangian $\sum_i p_i^\alpha - \mu (\sum_i p_i -1)$ becomes zero only at the maximum where $p_i = p_j \, \forall i,j$. The concavity of the Rényi entropy implies that the minimum is taken at the boundary where $p_1 = 1-p - (D-1)h$ and $p_2 = ... = p_D = h$, such that
\begin{align}
    \sum_{i=1}^{2^n} p_i^\alpha \geq (1 - p - ( D - 1 ) h)^\alpha + ( D - 1 ) h^\alpha + \frac{p}{h} h^\alpha,
    \label{eq:min_entropy}
\end{align}
using $p_i\leq h$ for all $i>D$ and therefore $\sum_{i>D} p_i^{\alpha - 1 + 1} \geq \frac{p}{h} h^\alpha$. The normalization constraint and $p_i \geq h$ for all $i \leq D$ further implies $Dh \leq 1-p$, which we can use to make the estimate 
\begin{align}
    \sum_i p_i^\alpha &\geq D h^\alpha + p h^{\alpha-1} \geq D^{1-\alpha} p^\alpha  \left( \left( \frac{1-\alpha}{\alpha} \right)^\alpha + \left( \frac{1-\alpha}{\alpha} \right)^{\alpha - 1} \right) = \frac{D^{1-\alpha} p^\alpha}{\alpha^\alpha (1-\alpha)^{1-\alpha}}.
\end{align}
The second inequality represents the global minimum of $D h^\alpha + p h^{\alpha-1}$ as a function of $h$. Plugging in $\sum_i p_i^\alpha = e^{S_{\mathcal{A}, \alpha} (1-\alpha)}$ yields the result.
\hfill \qedsymbol

\subsection{Approximation Error Bound -- Proof of Lemma \ref{lem:error}}
\label{app:error}
We show the error bound $\norm{ \ket{\psi} - U^\dagger \ket{0} } \leq \epsilon + L\sqrt{2(n-1)p}$ by separating the error into two parts. The first part is the accuracy of the optimizer $\epsilon = \norm{\ket{\psi} - \left( \prod_{i=L}^1 \mathcal{T}_D \circ G^\dagger_i \right) \ket{0} }$, which can be efficiently calculated classically. The second part is the truncation error arising for the sake of a bounded bond dimension. By the triangle inequality, we can extract $\epsilon$ via 
\begin{align}
    &\norm{ \ket{\psi} - \prod_{i=L}^1 G_i^\dagger \ket{0} } \leq \epsilon + \norm{\left( \prod_{i=L}^1 \left( \mathcal{T}_D \circ G^\dagger_i \right) - \prod_{i=L}^1 G^\dagger_i \right) \ket{0}}.
\end{align}
We show that every layer in the second term contributes at most linearly to the worst-case error. To do this, we prove the following estimation by induction
\begin{align}
    &\norm{\left( \prod_{i=L}^1 \left( \mathcal{T}_D \circ G^\dagger_i \right) - \prod_{i=L}^1 G^\dagger_i \right) \ket{\phi} } \leq \sum_{i=1}^L \norm{ ( \mathcal{T}_D \circ G^\dagger_i - G_i^\dagger ) \prod_{j = L}^{i+1} \left( \mathcal{T}_D \circ G^\dagger_j \right) \ket{\phi}} \qquad \forall \ket{\phi}.
    \label{eq:trunc_error}
\end{align}
The statement is trivially correct for $L=1$ and the induction step is done via
\begin{align}
    &\norm{\left( \prod_{i=L+1}^1 \left( \mathcal{T}_D \circ G^\dagger_i \right) - \prod_{i=L+1}^1 G^\dagger_i \right) \ket{\phi} } \nonumber \\
    &= \norm{\left( \prod_{i=L+1}^1 \left( \mathcal{T}_D \circ G^\dagger_i \right) - \left( \prod_{i=L}^1 G^\dagger_i \right) \left( \mathcal{T}_D \circ G^\dagger_{L+1} \right) + \left( \prod_{i=L}^1 G^\dagger_i \right) \left( \mathcal{T}_D \circ G^\dagger_{L+1} \right) - \prod_{i=L+1}^1 G^\dagger_i \right) \ket{\phi} } \nonumber \\
    &\leq \sum_{i=1}^L \norm{ \left( \mathcal{T}_D \circ G^\dagger_i  - G^\dagger_i \right) \left( \prod_{j =L}^{i+1} \left( \mathcal{T}_D \circ G^\dagger_j \right) \right) \left( \mathcal{T}_D \circ G^\dagger_{L+1} \right) \ket{\phi} } + \norm{ \left( \mathcal{T}_D \circ G^\dagger_{L+1} - G^\dagger_{L+1} \right) \ket{\phi} },
\end{align}
where we used the triangle inequality in the last step together with the induction assumption on the state $\left( \mathcal{T}_D \circ G^\dagger_{L+1} \right) \ket{\phi}$ for the first term and unitary invariance of the euclidean norm for the second. It has been shown \cite{Verstraete_2006, Schuch_2008} that the truncation error of a state $\ket{\psi}$ to a bond dimension $D$ state $\mathcal{T}_D \ket{\psi}$ is bounded by $\norm{\ket{\psi} - \mathcal{T}_D \ket{\psi}}^2 \leq 2 \sum_k \sum_{i=D+1}^{2^n} (\lambda_i^{[k]})^2 =: 2\sum_k p^{[k]}$ with $\lambda_i^{[k]}$ being the $i^{\text{th}}$ Schmidt value on the $k^\text{th}$ bond. We can estimate the right hand side of Eq.~\eqref{eq:trunc_error} by the truncation errors of the bond dimension $D$ MPS $\ket{\phi_i} := \prod_{j=L}^{i+1} (\mathcal{T}_D \circ G^\dagger_j) \ket{\psi}$. This reduces to $L$ truncation errors, which are in the worst case equal to $\sqrt{2(n-1)p}$. 

\hfill \qedsymbol

Although, we can find cases in which the triangle inequalities and the bounds by maximizing all tail Schmidt values are being exhausted, we expect the ``typical error'' of the state preparation to be below the derived bound.

\subsection{Absence of Barren Plateaus in CVD -- Proof of Lemma \ref{lem:barren}}
\label{app:barren}
At $\theta=0$, the gradient from Eq.~\eqref{eq:gradient} has a relatively simple form and can be calculated in graphical notation to be
\begin{align}
    &-i \Tr_\mathcal{A} \left( \Tr_{\mathcal{A}^c}\left( \ket{\psi} \bra{\psi} \right) \Tr_{\mathcal{A}^c}\left( \comm{\sigma_j^{(l)} \otimes \sigma_j^{(r)}}{\ket{\psi} \bra{\psi}} \right) \right) \nonumber \\[0.5cm]
    &= \vcenter{\hbox{\costgradpreI}} \quad - \quad \vcenter{\hbox{\costgradpreII}} \quad = \nonumber \\[0.5cm]
    &= \vcenter{\hbox{\costgradpostI}} \quad - \quad \vcenter{\hbox{\costgradpostII}} = 2i \Im\left( \vcenter{\hbox{\costgradpostI}} \right),
    \label{eq:grad_calc}
\end{align}
using the isometry property of $\Gamma_r \Lambda_r$ and $\Lambda_l \Gamma_l$. It becomes evident that for a product state, i.e. $\Lambda_0 = 1$, the gradient vanishes. As the Rényi entropy is always non-negative, product states are global minima of the cost function. The gradient also vanishes exactly, whenever the Schmidt values are evenly distributed, $\Lambda_0 \propto \mathds{1}$, or when all transition amplitudes between Schmidt vectors over $\sigma_l \otimes \sigma_r$ are real. In these cases, we are facing an unstable fixed point. 

Consider the example of two qubits in the state $\ket{\psi} = \alpha \ket{00} + \beta e^{i \phi} \ket{11}$. The gradient for the 2-Rényi entanglement entropy reads
\begin{align}
    -i \Tr\left( \rho \partial_{j} \rho \right) = 2 \Im ( &e^{-i \phi} \alpha \beta^3 \bra{1}\sigma_l\ket{0}\bra{1}\sigma_r\ket{0} \nonumber \\
    + &e^{i \phi} \alpha^3 \beta \bra{0}\sigma_l\ket{1}\bra{0}\sigma_r\ket{1} ),
\end{align}
with two non-vanishing elements for $\sigma_l = \sigma_r = X$: $-i \Tr\left( \rho \partial_{XX} \rho \right) = 2\sin(\phi) (\beta\alpha^3 - \alpha \beta^3)$ and similarly $\sigma_l = \sigma_r = Y$. One can quickly see that the gradient vanishes for the product state $\alpha = 0$ or $\beta = 0$, as well as for the Bell state $\alpha = \beta$. As the Bell state is maximally entangled, it can only be a maximum of the entanglement entropy and thus an unstable fixed point. Indeed, adding a slight imbalance in the amplitudes $\alpha^2 \to \frac{1}{2} + \delta, \beta^2 \to \frac{1}{2} - \delta$ results in an increasing gradient. Another case, in which the gradient vanishes, is $\phi=0$, where all transition amplitudes with respect to $\sigma_l \otimes \sigma_r$ are real-valued. Also here, the gradient increases whenever $\phi \to \pm \delta$ is perturbed around the fixed point $\phi = 0$. the gradient then pulls toward $\alpha = 1$ or $\beta = 1$ depending on the sign of $\phi$. 

More generally, a state $\ket{\psi} = \sum_{j} \lambda_j \ket{\phi_j^{(l)}} \otimes \ket{\phi_j^{(r)}}$ with only real transitions $\bra{\phi^{(l)}_j} \sigma^{(l)} \ket{\phi^{(l)}_k}\bra{\phi^{(r)}_j} \sigma^{(r)} \ket{\phi^{(r)}_k}$, will have a vanishing gradient. However, that fixed point is easily escaped by tilting the relative phase towards one of the $\ket{\phi_j^{(r)}}$, for instance. 

We now prove Lemma \ref{lem:barren} using the result of Eq.~\eqref{eq:grad_calc}. To do so, we consider uniformly distributed isometries $A = \Lambda_l \Gamma_l$ and $B = \Gamma_r \Lambda_r$ and calculate the first and second moment of the gradient $ \partial_i S_{\mathcal{A}, 2} \Big\vert_{\theta = 0}$. Using the isometric property of $A$ and $B$, we can formulate the calculation in terms of unitaries $U, V$ defined via
\begin{align}
    \vcenter{\hbox{\isometry{$A$}{$U$}}} \hspace{1cm} \vcenter{\hbox{\isometry{$B$}{$V$}}},
\end{align} 
and consider them distributed via the Haar measure on SU(2$D$). With this, we can integrate polynomials of $U$ and $U^\dagger$ or $V$ and $V^\dagger$, respectively, using Weingarten calculus, in particular we make use of the formulas
\begin{align}
    \int_{\text{SU(2$D$)}} dU \, U_{ab} \overline{U_{cd}} &= \frac{\delta_{ac} \delta_{bd}}{2D} \label{eq:WeingartenI}\\
    \int_{\text{SU(2$D$)}} dU \, U_{ab} U_{cd} \overline{U_{ef}} \, \overline{U_{gh}} &= \frac{1}{4D^2-1} (\delta_{ae}\delta_{bf}\delta_{cg}\delta_{dh} + \delta_{ag}\delta_{bh}\delta_{ce}\delta_{df}) - \frac{1}{2D(4D^2 - 1)} (\delta_{ae}\delta_{bh}\delta_{cg}\delta_{df} + \delta_{ag}\delta_{bf}\delta_{ce}\delta_{dh}),
    \label{eq:WeingartenII}
\end{align}
which also can be written in graphical notation
\begin{align}
    \int_{\text{SU(2$D$)}} dU \, \vcenter{\hbox{\unitary{$U$}{$a$}{$b$} \unitary{$\overline{U}$}{$c$}{$d$}}} &= \frac{1}{2D} \vcenter{\hbox{\deltaTensor{$a$}{$b$}{$c$}{$d$}}} \label{eq:WeingartenIII} \\
    \int_{\text{SU(2$D$)}} dU \, \vcenter{\hbox{\unitary{$U$}{$a$}{$b$} \unitary{$U$}{$c$}{$d$} \unitary{$\overline{U}$}{$e$}{$f$}\unitary{$\overline{U}$}{$g$}{$h$}}} &= \frac{1}{4D^2 - 1} \left( \vcenter{\hbox{\deltaTensorII{$a$}{$b$}{$c$}{$d$}{$e$}{$f$}{$g$}{$h$}}} + \vcenter{\hbox{\deltaTensorIII{$a$}{$b$}{$c$}{$d$}{$e$}{$f$}{$g$}{$h$}}} \right) \nonumber \\
    &- \frac{1}{2D(4D^2-1)} \left( \vcenter{\hbox{\deltaTensorIV{$a$}{$b$}{$c$}{$d$}{$e$}{$f$}{$g$}{$h$}}} + \vcenter{\hbox{\deltaTensorV{$a$}{$b$}{$c$}{$d$}{$e$}{$f$}{$g$}{$h$}}} \right).
    \label{eq:WeingartenIV}
\end{align}
Being ultimately a trace of Pauli operators, the expected value of the gradient thus vanishes. The first term of Eq.~\eqref{eq:grad_calc} reads for instance
\begin{align}
    \mathds{E}_A \mathds{E}_B \left[ \vcenter{\hbox{\gradAB}} \right] = \int_{\text{SU(2$D$)}} dU \, \int_{\text{SU(2$D$)}} dV \, \left[ \vcenter{\hbox{\gradUV}} \right] = \frac{1}{4} \Tr(\sigma_l) \Tr(\sigma_r) \Tr(\Lambda_0) \Tr(\Lambda_0^3) = 0,
    \label{eq:first_moment}
\end{align}
and similarly the second term vanishes. The second moment is calculated using Eq.~\eqref{eq:WeingartenII}, whereas two of the four contractions vanish as they separate into terms of the form of Eq.~\eqref{eq:first_moment}. The remaining terms are integrals over squares of $\vcenter{\hbox{\resizebox{0.72cm}{0.6cm}{\gradAB}}}$, once with $\Lambda_0^3$ on the upper contraction and once on the lower contraction, as well as the mixed term between both. The squared terms read
\begin{align}
    \mathds{E}_A \mathds{E}_B \left[ \begin{matrix}
        \vcenter{\hbox{\gradAB}} \\
        \vcenter{\hbox{\gradAB}}
    \end{matrix} \right] &= \int_{\text{SU(2$D$)}} dU \, \int_{\text{SU(2$D$)}} dV \, \left[ \begin{matrix}
        \vcenter{\hbox{\gradUV}} \\
        \vcenter{\hbox{\gradUV}}
    \end{matrix} \right] = \frac{1}{2D+1} \int_{\text{SU(2$D$)}} dV \vcenter{\hbox{\secondmomentstepI}} \nonumber \\
    &= \frac{\Tr\left( \Lambda_0^4 \right)^2}{(2D+1)^2}
\end{align}
The second squared term yields the same outcome and the mixed term can be calculated similarly by just swapping $\Lambda_0$ and $\Lambda_0^3$ in one of the terms. This yields $\frac{\Tr\left( \Lambda_0^6 \right)}{(2D+1)^2}$ instead. Plugging in everything into the formula of the gradient of the Rényi entropy, Eq.~\eqref{eq:gradient}, finally yields the second moment as claimed.
\begin{align}
    \mathds{E}_{A} \mathds{E}_{B} \left[ \left( \partial_i S_{\mathcal{A}, 2} \Big\vert_{\theta = 0} \right)^2 \right] &= \left(-\frac{2}{\Tr(\Lambda_0^4)} \right)^2 \frac{2}{(2D+1)^2} \left( \Tr(\Lambda_0^4)^2 - \Tr(\Lambda_0^6) \right) = \frac{8}{(2D+1)^2} \left( 1 - \frac{\Tr(\Lambda_0^6)}{\Tr(\Lambda_0^4)^2} \right). \qed
\end{align}

\section{CVD on the GHZ, Cluster and AKLT state}
\label{app:simples}
\begin{figure}
    \centering
    \begin{tabular*}{\textwidth}{c@{\extracolsep{\fill}}cc}
        \textbf{(a)} \hspace{1.25cm} \textbf{GHZ state} & \textbf{Cluster State} & \textbf{AKLT State} \hspace{1.7cm} \\
        \hspace{1.8cm} \scriptsize{$S_\infty = - \log(\lambda_1^2)$} & \scriptsize{$S_\infty = - \log(\lambda_1^2)$} & \scriptsize{$S_\infty = - \log(\lambda_1^2)$} \hspace{1.6cm} 
    \end{tabular*}
    $\vcenter{\hbox{\includegraphics[width=0.32\linewidth]{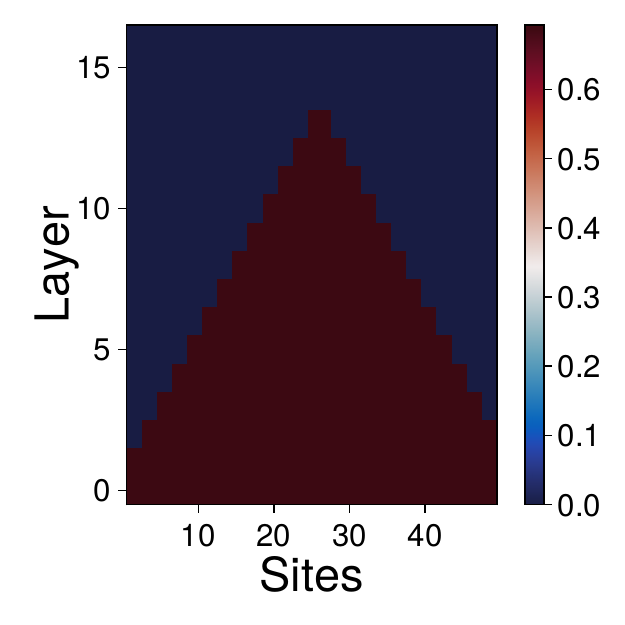}}}$
    $\vcenter{\hbox{\includegraphics[width=0.32\linewidth]{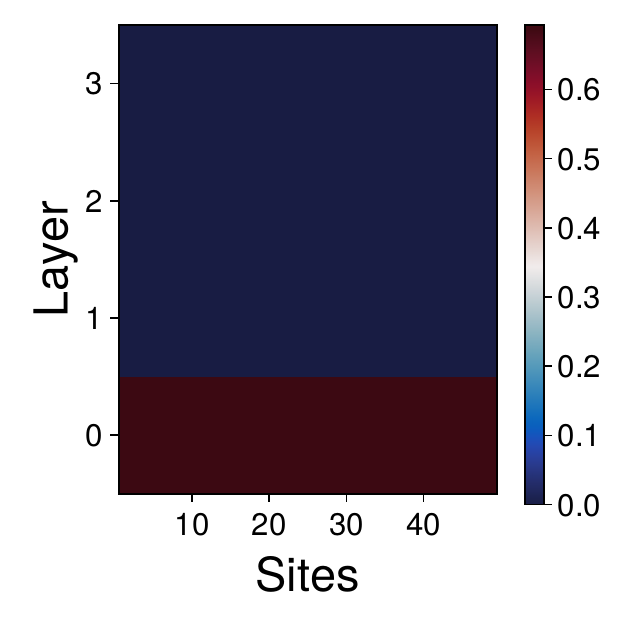}}}$
    $\vcenter{\hbox{\includegraphics[width=0.32\linewidth]{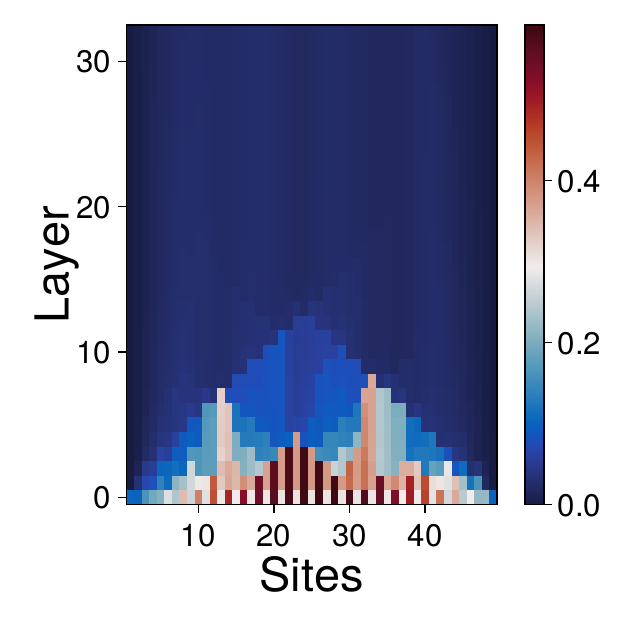}}}$
    
    \raggedright \textbf{(b)} \hspace{8cm} \textbf{(c)} 

    $\vcenter{\hbox{\includegraphics[width=0.45\linewidth]{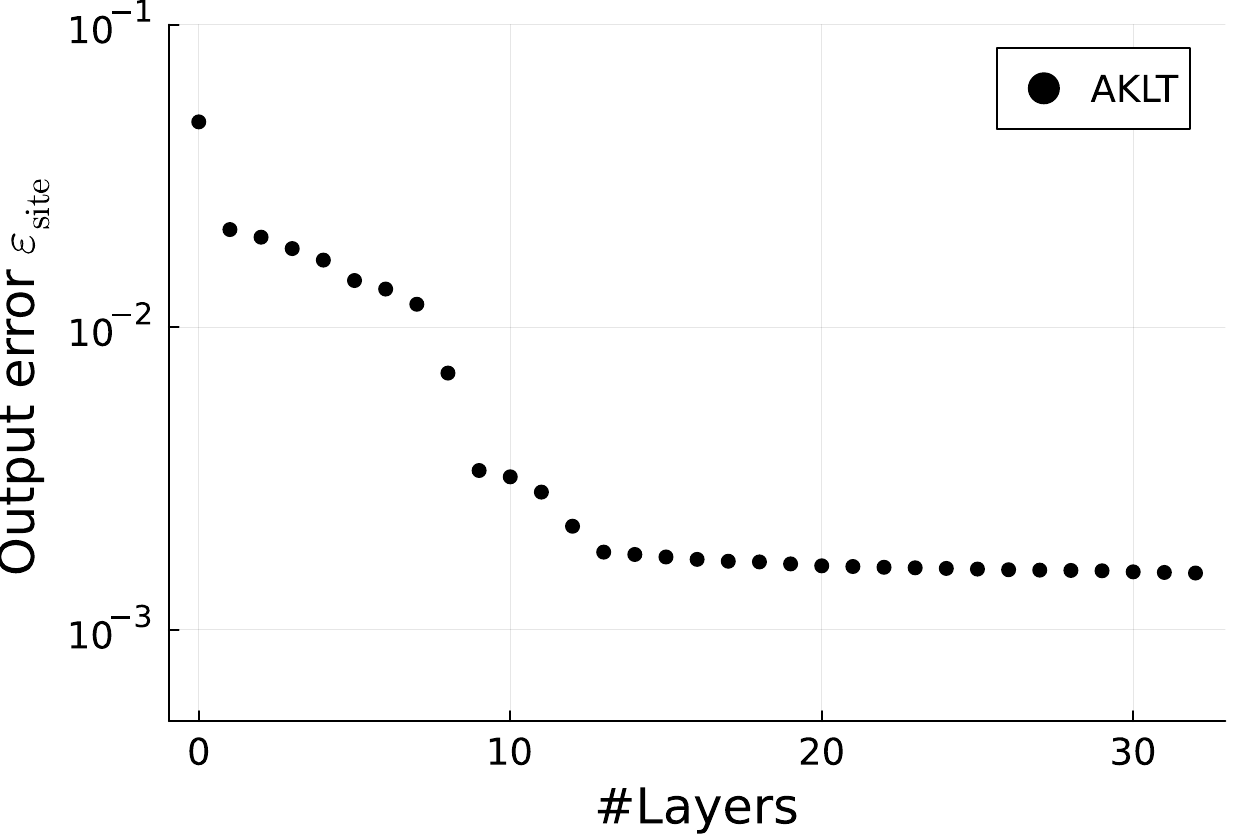}}}$
    \hspace{1cm}
    $\vcenter{\hbox{\includegraphics[width=0.45\linewidth]{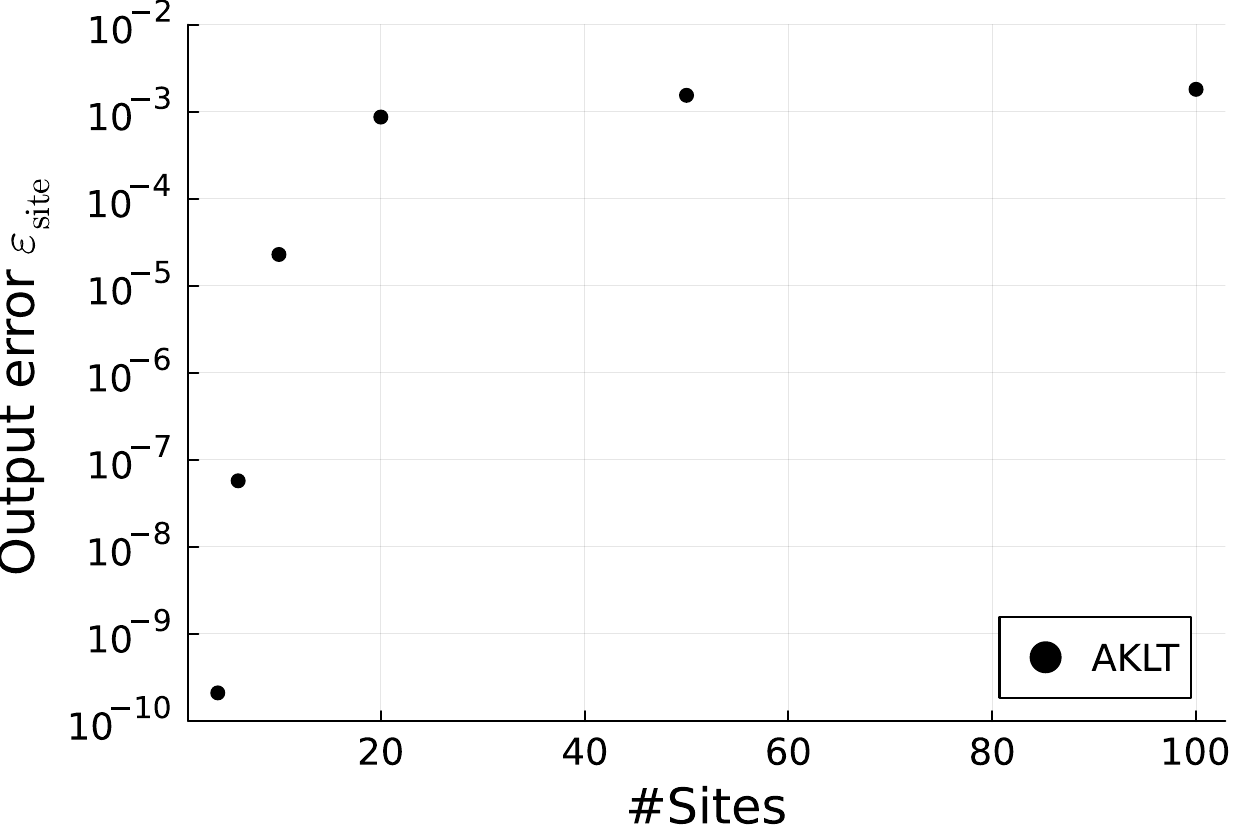}}}$
    \caption{Disentanglement of the GHZ, Cluster and AKLT state on 50 sites using $\alpha = 1$. (a) A plot of the tail weights reveals that exact preparation circuits for the GHZ and the cluster state can be found with the expected circuit depth, that is $\frac{n}{2}$ for the GHZ state and 2 for the cluster state. (b) The AKLT state on 25 spin-1 sites (50 qubit sites restricted to the triplet subspaces) is approximately prepared in 10 -- 13 layers, as the overlap error comes close to $\epsilon_{\rm site} = 10^{-3}$. After that, some tail weights can be exactly set to 0 as can be seen in the third color plot in (a). (c) The AKLT state preparation is scalable to larger system sizes, as $\epsilon_{\rm site}$ saturates.}
    \label{fig:simples}
\end{figure}
Some matrix product states are of conceptual interest in quantum information theory and quantum many body physics. In the following, we want to consider the GHZ state $\ket{\psi} = \frac{1}{\sqrt{2}} \left( \bigotimes_{i=1}^n \ket{0} + \bigotimes_{i=1}^n \ket{1} \right)$, the cluster state, which is the ground state of the cluster Ising Hamiltonian $H = \sum_i Z_{i-1} X_i Z_{i+1}$ and the (unnormalized) AKLT state,
\begin{align}
    \bigotimes_{\rm odd} P^{(1)} \bigotimes_{\rm even} \ket{\psi^-} = \vcenter{\hbox{\AKLT}},
    \label{eq:AKLT}
\end{align}
which is built from neighboring singlet states $\ket{\psi^-}$ with each an endpoint of two different singlets projected onto the triplet subspace. In the literature, one often finds the pictorial representation shown in Eq.~\eqref{eq:AKLT} of the AKLT state, where the dots are spin 1/2 sites, straight lines denote singlet states and the ellipse depicts a projection onto the triplet subspace spanned by $\{\ket{00}, \ket{11}, \frac{1}{\sqrt{2}} (\ket{01} + \ket{10})\}$. 

Our results to these three states are illustrated in Fig.~\ref{fig:simples}. It is well-known that the GHZ state exhibits topological order and therefore cannot be prepared with a local circuit of sublinear depth. The cluster state, on the other hand, is known for having particularly low state complexity being exactly preparable with a depth 2 circuit. Our results are in agreement with these expectations and show exact preparation of the GHZ state in circuit depth $\frac{n}{2}$ and the cluster state with depth 2.

The AKLT state does not have a straightforward circuit representation except the trivial approach of sequential preparation. There has been some effort to find preparation circuits \cite{Chen_2023, Chen_2024} with imaginary time evolution on the AKLT Hamiltonian. We manage to find a shallow quantum circuit to prepare the AKLT state on 25 spin-1 sites, where each spin-1 site is encoded into the triplet subspace of two qubits. We reach a per-site overlap error of $1.8 \times 10^{-3}$ with 13 layers of disentangling gates. Although, this is a non-negligible remaining error, it is below the typical gate error rate of near-term quantum hardware. Better fidelities can be achieved using a lower cutoff for the Schmidt values and deeper circuits.

\end{document}